\documentclass[twocolumn,prb,color,psfig,superscriptaddress]{revtex4}
\usepackage{graphicx}
\usepackage{epsfig}

\def\beq{\begin{equation}}
\def\eeq{\end{equation}}

\begin{document}
\title{Interplay of \emph{p-d} and \emph{d-d} charge transfer transitions in rare-earth perovskite manganites}
\author{A.S. Moskvin}
\affiliation{Ural State University, 620083 Ekaterinburg,  Russia}
\author{A.A. Makhnev}
\affiliation{Institute of Metal Physics, 620219 Ekaterinburg, Russia}
\author{L.V. Nomerovannaya}
\affiliation{Institute of Metal Physics, 620219 Ekaterinburg, Russia}
\author{N.N. Loshkareva}
\affiliation{Institute of Metal Physics, 620219 Ekaterinburg, Russia}
\author{A.M. Balbashov}
\affiliation{Moscow Power Engineering Institute, 105835 Moscow, Russia}
\date{\today}

\begin{abstract}
We have performed both theoretical and experimental study of optical response of parent perovskite manganites RMnO$_3$ with a main goal to elucidate nature of clearly visible optical features. Starting with a simple cluster model approach  we addressed the both  one-center (\emph{p-d}) and two-center (\emph{d-d}) charge transfer (CT) transitions, their polarization properties, the role played by structural parameters, orbital mixing, and spin degree of freedom. Optical complex dielectric function of single crystalline samples of RMnO$_3$ (R=La, Pr, Nd, Sm, Eu) was measured by ellipsometric technique at room temperature in the spectral range from 1.0 to 5.0 eV for two light polarizations: $\bf E \parallel \bf c$ and $\bf E \perp \bf c$. The comparative analysis of the spectral behavior of  $\varepsilon _1$ and  $\varepsilon _2$ is believed to provide a more reliable assignment of spectral features. We have found an overall agreement between experimental spectra and theoretical predictions based on the theory of one-center \emph{p-d} CT transitions  and inter-site \emph{d-d} CT transitions. Our experimental data and theoretical analysis evidence a dual nature of the dielectric gap in nominally stoichiometric matrix of perovskite manganites RMnO$_3$, it is formed by a superposition of forbidden or weak dipole allowed \emph{p-d} CT transitions and inter-site \emph{d-d} CT transitions. In fact, the parent perovskite manganites RMnO$_3$ should rather be sorted neither into the CT insulator nor the Mott-Hubbard insulator in the Zaanen, Sawatzky, Allen  scheme.  
\end{abstract}

\maketitle

\section{Introduction}

The nature of the low-energy optical electron-hole excitations in the
insulating transition metal 3\emph{d} oxides represents one of the most important
challenging issues for these strongly correlated systems. 
All these excitations are
especially interesting because they could play a central role in multiband Hubbard
models used to describe both the insulating state
and the unconventional states developed under electron or hole doping. Because of the matrix element effect the optical response does provide only an indirect information about the density of states. Nevertheless it remains one of the most efficient technique to inspect the electronic structure and energy spectrum.

It is now believed that  the most intensive low-energy electron-hole excitations
 in insulating 3\emph{d} oxides correspond to the charge transfer (CT) transitions while different phonon-assisted crystal field transitions are generally much weaker. Namely the CT transitions are considered as a likely source of the optical and magnetooptical response of the 3\emph{d} metal-based oxide compounds in a wide spectral range 1-10 eV, in particular, the fundamental absorption edge. The low-energy dipole-forbidden \emph{d-d} orbital excitations (or crystal field transitions) are characterized by the oscillator strengths which are smaller by a factor
$10^2-10^3$ than that for the dipole-allowed \emph{p-d} CT transitions and usually correspond to contributions 
to the dielectric function $\varepsilon_2$ of the order of 0.001-0.01.

Despite  CT transitions are well established concept in the solid state physics, 
 their theoretical treatment remains  rather naive and did hardly progress during last decades. Usually it is based on the {\it one-electron} approach with some 2\emph{p}-3\emph{d}    or, at best, 2\emph{p}$\rightarrow$ 3\emph{d}$\,t_{2\mathrm{g}}$, 2\emph{p}$\rightarrow$3\emph{d}$\,e_{\mathrm{g}}$ CT transitions in 3\emph{d} oxides. In terms of the Hubbard model, this is a CT transition from the nonbonding oxygen band to the upper Hubbard band.
But such a simplified approach to CT states and transitions in many cases appears to be absolutely insufficient and misleading even for qualitatively explaining the observed optical and magnetooptical properties. First, one should generalize the concept of CT transitions taking into account  the conventional transition between the lower and upper Hubbard bands which corresponds to an inter-site \emph{d-d} CT  transition (intersite transition across
the Mott gap).

Several important problems  are hardly addressed in the  current analysis of optical spectra, including the relative role of different initial and final orbital states and respective CT channels, strong intra-atomic correlations, effects of strong electron and lattice relaxation for CT states, the transition matrix elements, or  transition probabilities, probable change in crystal fields and correlation parameters accompanying the charge transfer.

 One of the central issues in the analysis of electron-hole
excitations  is whether low-lying states   are comprised of free
charge carriers or excitons. A conventional approach implies that if
the Coulomb interaction is effectively screened and weak, then the
electrons and holes are only weakly bound and move essentially
independently as free charge-carriers. However, if the Coulomb
interaction between electron and hole is strong, excitons are believed to
form, i.e.\ bound particle-hole pairs with strong correlation of
their mutual motion.

Our work was stimulated by the lack of  detailed and reliable studies of electron-hole excitations and of a proper understanding of the relative role of \emph{p-d} and \emph{d-d} CT transitions in rare-earth perovskite manganites RMnO$_3$ (R= rare earth or yttrium) to be parent systems for the colossal magnetoresistive\,\cite{Tokura} and multiferroic\,\cite{Kimura} materials. Given the complex phase diagram of
this class of materials, studies of the nominally stoichiometric parent
compound could give an insight into  physics governing the
doped version of these manganite oxides.

Virtually all the experimental optical data available for manganites are focused on the LaMnO$_3$. 
Except for paper by Kim {\it et al.}\cite{Kim2} on the absorption spectroscopy of thin films of RMnO$_3$ (R=La, Pr, Nd, Gd, Tb), there has been no effort to perform a comparative study of the optical response for different rare-earth perovskite manganites. 
The optical conductivity spectrum of
LaMnO$_3$ exhibits two broad intensive bands centered around $2.0$ and $4$-$5$ eV.
\cite{Arima,Okimoto,Jung,Takenaka} However, it has remained unclear just what
the nature of the  related electron-hole excitations. Some authors \cite{Arima,Okimoto,Takenaka} assign these both
 features  to the dipole-allowed \emph{p-d} CT transitions like
   $t_{2g}^{3}e_{g}^{1}-t_{2g}^{3}e_{g}^{2}\underline{L}$ and
   $t_{2g}^{3}e_{g}^{1}-t_{2g}^{4}e_{g}^{1}\underline{L}$ ($\underline{L}$
   denoting a ligand hole), respectively. However, others \cite{Jung,Allen}
assign the low-energy peak to the Jahn-Teller orbiton-excitation ${}^{5}E_g-{}^{5}E_{g}^{'}$
transition, or doubly-forbidden (parity and orbital quasimomentum) \emph{d-d}-like
crystal-field transition between two ${}^{5}E_g$-sublevels separated by a
splitting due to a low-symmetry crystalline field.
 Both interpretations being particularly qualitative suffer from many
shortcomings and give rise to many questions concerning the details of the
charge transfer states or expected extremely weak intensity for the \emph{d-d} crystal field
transitions.
Pronounced temperature rearrangement of the optical spectral weight both for low- and high-energy bands was uncovered by  Quijada {\it et al.}\cite{Quijada}. 
These authors were seemingly the first who made a valid conclusion that the dominant contribution to the optical spectral weight of the conductivity peak at 2.0 eV is provided by the CT hopping between nearest-neighbor manganese ions, or inter-site \emph{d-d} CT transitions. 

Until recently main body of the optical data for manganites was  obtained  by reflectivity measurements 
followed by a Kramers-Kr\"{o}nig (KK) transformation usually accompanied by a
number of unavoidable uncertainties as regards the peak positions
and intensities of weak spectral features. It often implies a parasitic contribution due to a deterioration of the sample surface \cite{Okimoto}, that can give rise to some ambiguities due to problems  with KK transformation. The technique of ellipsometry provides significant advantages over conventional reflection methods in that it is self-normalizing and does not require reference measurements, and optical complex dielectric function
$\varepsilon=\varepsilon_{1}-i\varepsilon_{2}$  is obtained directly
without a Kramers-Kr\"onig transformation. The comparative analysis of the spectral behavior of $\varepsilon_{1}$ and $\varepsilon_{2}$ is believed to provide a more reliable assignement of spectral features.

First ellipsometric measurements for the single crystalline LaMnO$_3$ samples were performed by Loshkareva {\it et al.}\cite{optics-PS} at room temperature and for the spectral range 1.0-5.0 eV.
 Later on the ellipsometric measurements were performed for an untwinned crystal of LaMnO$_3$ by Kovaleva {\it et al.}\,\cite{Kovaleva,Kovaleva-09} in a wide temperature range. The authors have presented a detailed quantitative analysis of the pronounced redistribution of the spectral weight (SW) near the N\'eel temperature.  They concluded  that the low-energy optical band around 2 eV consists of
three distinct bands all assigned to intersite \emph{d-d} CT transitions, and that LaMnO$_3$ is a Mott-Hubbard  rather than a charge transfer \emph{p-d} insulator as argued earlier (see e.g., Refs.\onlinecite{Tobe,Moskvin-Mn}). A similar interpretation of spectral features near 2 eV observed in multiferroic TbMnO$_3$ was reported very recently by Bastjan {\it et al.}\,\cite{Bastjan}. Many researchers\,\cite{Tobe,Kovaleva,Kovaleva-09,Lawler,Kim1} pointed to a fine sub-peak structure of the  low-energy 2 eV band. Early optical transmission spectra of the LaMnO$_3$ films\,\cite{Lawler} revealed a fine structure  with
 two  features near $1.7$ and $2.4$ eV, respectively, which were
   assigned to the Mn$^{3+}$ \emph{d-d} crystal-field transition ${}^{5}E_{g}-{}^{3}T_{1g}$,
  split by the JT effect.  Up to this point, there has been little effort to understand this mysterious sub-peak structure. First theoretical cluster model analysis of the \emph{p-d} CT transitions in the MnO$_6$ octahedra\,\cite{Moskvin-Mn} did predict several low-energy forbidden or relatively weak allowed electro-dipole transitions with an onset at 1.7 eV to be a "precursor" of a strong dipole-allowed \emph{p-d} CT transition at 4.5-4.7 eV.

Such an ambiguity leaves  the question of the nature of the main optical
transitions and low-lying electron states in LaMnO$_3$  far from being resolved. 
Furthermore, it seems that there is a missing qualitative aspect of the problem that so far escaped identification. 

The one-electron band models, including such modern modifications as LDA$+U$, do not elucidate the matter, first of all since they  fail to reproduce important intra-atomic correlation effects which form the term structure both for the ground and excited CT configurations.\cite{Ahn,Ravindran} In this connection we would like to emphasize the priority of the cluster model of the CT transitions which
showed itself to advantage in explaining the optical and magnetooptical spectra
of orthoferrites, iron garnets, and a number of dielectric cuprates and
manganites (see, e.g., Refs.\onlinecite{Kahn,Moskvin-Ferro,Moskvin-Mn,Moskvin1,Moskvin2,ferrites}). The model is marked by the physical
clearness, possibility of detailed  account for the electron correlation
and the crystal field effects as well.

Our efforts were focused on the theoretical and experimental studies of the low energy CT bands peaked near 2 eV with a particular interest in the fine-structure effects. To this end  we first address a theoretical analysis of a large variety of  different CT states and  CT transitions in the perovskite  manganites based on the distorted MnO$_6$ octahedra being a basic element of their crystalline and electronic structure. Then we present the results of optical ellipsometry for a number of manganites RMnO$_3$ (R= Pr, Nd, Sm, Eu) with a supposedly reduced spectral weight of the \emph{d-d} CT transitions and a better manifestation of the fine-structure effects.   

The rest of the paper is organized as follows.  In Sec.\ II we
shortly address the electronic structure, the energy spectrum of
a MnO$_6$ cluster, and the one-center \emph{p-d} CT transitions.
Two-center \emph{d-d} CT transitions  are considered in Sec.\ III. The
experimental results of ellipsometric optical measurements for perovskite manganites RMnO$_3$ (R=La, Pr, Nd, Sm, Eu) are presented in Sec.\ IV. In Sec.\ V we  address an extended discussion of experimental data which includes a Lorentzian fitting of dielectric function,  and a short overview of electronic structure for parent perovskite manganites as derived from optical data.

\section{One-center (\lowercase{\emph{p-d}}) charge transfer transitions in orthorhombic manganites}

\subsection{Electronic structure of octahedral 3\emph{d}-metal-oxygen
MeO$_6$ centers in perovskites}

The electronic states in  strongly correlated 3\emph{d} oxides manifest
both significant correlations and dispersional features. The
dilemma posed by such a combination is the overwhelming number of
configurations which must be considered in treating strong
correlations in a truly bulk system. One strategy to deal with
this dilemma is to restrict oneself to small clusters, creating
model Hamiltonians whose spectra may reasonably well represent the
energy and dispersion of the important excitations of the full
problem. Naturally, such an approach has a number of principal
shortcomings, including the boundary conditions, the breaking of
local symmetry of boundary atoms, and so on.
Nevertheless, this method provides a clear physical
picture of the complex electronic structure and the energy
spectrum, as well as the possibility of quantitative modelling.
In a certain sense the cluster
calculations might provide a better description of the overall
electronic structure of  insulating  3\emph{d} oxides  than the 
band structure calculations, mainly  due to a better account for correlation effects.

 Five Me 3\emph{d} and eighteen  oxygen O 2\emph{p} atomic
orbitals in octahedral MeO$_6$ complex with  the point symmetry group $O_h$
form both hybrid Me 3\emph{d}-O 2\emph{p}  bonding and antibonding $e_g$ and $t_{2g}$
molecular orbitals (MO), and purely oxygen nonbonding $a_{1g}(\sigma)$, $t_{1g}(\pi)$,
$t_{1u}(\sigma)$, $t_{1u}(\pi)$, $t_{2u}(\pi)$ orbitals.
 Nonbonding $t_{1u}(\sigma)$ and $t_{1u}(\pi)$ orbitals with
the same symmetry  are hybridized due to the oxygen-oxygen O 2\emph{p}$\pi$ - O
2\emph{p}$\pi$ transfer. The relative energy position of different nonbonding oxygen
orbitals is of primary importance for the spectroscopy of the oxygen-3\emph{d}-metal
charge transfer. This is firstly determined by the bare energy separation
$\Delta \epsilon _{2p\pi \sigma}=\epsilon _{2p\pi }-\epsilon _{2p\sigma}$
between O 2\emph{p}$\pi$ and O 2\emph{p}$\sigma$ electrons.
\begin{figure}[t]
\includegraphics[width=8.5cm,angle=0]{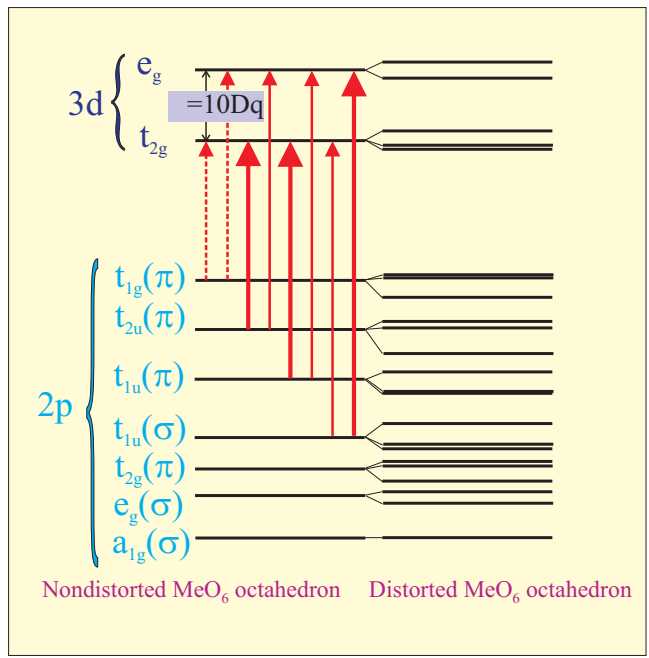}
\caption{(Color online) The diagram of Me 3\emph{d}-O 2\emph{p} molecular orbitals  for the MeO$_{6}$
octahedral center. The O 2\emph{p} - Me 3\emph{d} charge transfer transitions are shown by
arrows: strong dipole-allowed  $\sigma -\sigma$  and $\pi -\pi$ by thick solid
arrows;  weak dipole-allowed  $\pi -\sigma$ and $\sigma -\pi$ by thin solid
arrows; weak dipole-forbidden low-energy transitions by thin dashed arrows,
respectively.} \label{fig1}
\end{figure}
  Since the O 2\emph{p}$\sigma$ orbital points towards the two neighboring positive
   $3\emph{d}$ ions, an electron in this orbital has its energy lowered
   by the Madelung potential as compared with the O 2\emph{p}$\pi$ orbitals,
   which are oriented perpendicular
   to the respective 3\emph{d}-O-3\emph{d} axes. Thus, Coulomb arguments  favor
    the positive sign of the $\pi -\sigma$ separation
     $\epsilon _{p\pi}-\epsilon _{p\sigma}$ which
  numerical value   can be easily
   estimated in frames of the well-known point charge model, and appears to be of the order of
   $1.0$ eV.
   In a first approximation, all the $\gamma (\pi )$ states
    $t_{1g}(\pi),t_{1u}(\pi),t_{2u}(\pi)$ have the same energy. However,
the O 2\emph{p}$\pi$-O 2\emph{p}$\pi$ transfer yields the energy correction
   to bare energies with the largest value
   and positive  sign  for
   the $t_{1g}(\pi)$ state. The energy of the $t_{1u}(\pi)$ state drops due to
   a hybridization with the cation 4p$t_{1u}(\pi)$ state. In other words, the $t_{1g}(\pi)$ state is believed
   to be the highest in energy non-bonding oxygen state.
   For illustration, in Fig.1 we show the energy spectrum of the 3\emph{d}-2\emph{p} manifold in
   octahedral complexes like MeO$_6$ with the relative energy position of the levels
   according to the quantum chemical calculations \cite{Licht} for the
   FeO$_{6}^{9-}$ octahedral complex in a lattice environment typical for perovskites
   like LaFeO$_3$ and LaMnO$_3$. It should be emphasized one more that the top of the oxygen electron band is composed of O 2\emph{p}$\pi$ nonbonding orbitals that predetermines the role of oxygen states in many physical properties of 3\emph{d} perovskites.  
   

 {\it Conventional} electronic structure of octahedral MnO$_6$
complexes is related with the configuration of the completely filled O 2\emph{p} shells
and partly filled Mn 3\emph{d} shells. The typical high-spin ground state
configuration and crystalline term for Mn$^{3+}$ in octahedral crystal field or
for the octahedral MnO$_{6}^{9-}$ center is $t_{2g}^{3}e_{g}^1$ and
${}^{5}E_{g}$, respectively. Namely this orbital doublet results in a vibronic
coupling and Jahn-Teller (JT) effect for the MnO$_{6}^{9-}$ centers, and
cooperative JT ordering in LaMnO$_3$. In the framework of a crystal field model
the ${}^{5}E_{g}$ term originates from the  (3\emph{d}$^{4})\,{}^{5}D$ term of  free
Mn$^{3+}$ ion. 

{\it Unconventional} electronic configuration of octahedral MnO$_6$
complexes is related with a {\it charge transfer  state} with  one hole in the O
2\emph{p} shells\cite{Moskvin-02}. The excited CT configuration $\underline{\gamma}_{2p}^1$ 3\emph{d}$^{n+1}
\;$ arises from the transition of an electron from the MO predominantly anionic
in nature (the $\gamma_{2p}$ hole  in the core of the anionic MO being hereby
produced), into an empty 3\emph{d} type MO ($t_{2g}\,$ or $\,e_g$). The transition
between the ground configuration and the excited one can be presented as the
$\,\gamma_{2p}\,\rightarrow\,$ 3\emph{d}$(t_{2g},e_g)$ CT transition.

The CT configuration consists of two partly filled subshells, the ligand $\;
\gamma _{2p}\,$ and the cation 3\emph{d}$(t_{2g}^{n_{1}}e_{g}^{n_{2}})$ shell,
respectively. The latter configuration in the case of CT states in the
MnO$_{6}^{9-}$ center nominally corresponds to Mn$^{2+}$ ion.
It should be emphasized that  the oxygen hole having occupied the
{\it non-bonding} $\; \gamma _{2p}\,$ orbital does interact {\it ferromagnetically}
with the 3\emph{d}$(t_{2g}^{n_{1}}e_{g}^{n_{2}})$ shell. This rather strong 
ferromagnetic coupling results in Hund rule for the CT configurations with the high-spin ground states. The maximal value of the total spin for
the Hund-like CT state in MnO$_{6}^{9-}$ center equals $S=3$, that points to
some perspectives of unconventional magnetic signatures of these states.

\subsection{Electric dipole matrix elements}

\begin{figure}[t]
\includegraphics[width=8.5cm,angle=0]{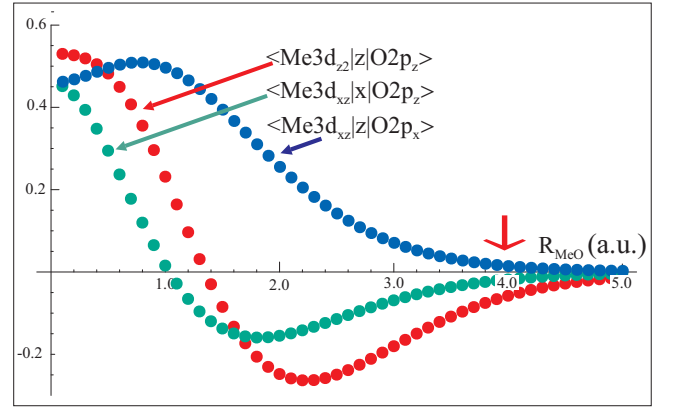}
\caption{(Color online) Two-site dipole matrix elements against Me\,3\emph{d}-O\,2\emph{p} separation. The arrow near 4 a.u. points to  typical Me-O separations. For illustration we choose both relatively large integrals $\langle 3d_{z^2}|z|2p_z\rangle$ governed by the  Me\,3\emph{d}-O\,2\emph{p} $\sigma$-bond and the relatively small ones $\langle 3d_{xz}|z|2p_x\rangle$ and $\langle 3d_{xz}|x|2p_z\rangle$ governed by the  Me\,3\emph{d}-O\,2\emph{p} $\pi$-bond. We make use of hydrogen-like radial wave functions with the Clementi-Raimondi effective charges\cite{Clementi-Raimondi} $Z_{O2p}^{eff}$=4.45  and $Z_{Me3d}^{eff}$=10.53 typical for Mn$^{3+}$  ion.} \label{fig2}
\end{figure}

The one-electron electric dipole matrix elements for MeO$_6$ octahedral center can be written with the aid of Wigner-Eckart theorem\cite{Varshalovich} as follows (see Ref.\onlinecite{Moskvin-Mn} for details)
\begin{equation}
\langle \gamma _{u}\mu|{\hat d}_{q}|\gamma _{g}\mu ^{'}\rangle =
(-1)^{j(\gamma _{u})-\mu}
\left\langle \begin{array}{ccc}
\gamma _{u} & t_{1u} & \gamma _{g} \\
-\mu & q & \mu ^{'}
\end{array}
\right\rangle ^{*} \langle \gamma _{u}\|\hat d\|\gamma _{g}\rangle \,,
\label{d}
\end{equation}
where $\left\langle
\begin{array}{ccc} \cdot & \cdot & \cdot \\ \cdot & \cdot & \cdot
\end{array}
\right\rangle $ is the Wigner coefficient for the cubic point group O$_h$, $j(\Gamma )$ the so-called quasimomentum number, $\langle \gamma _{u}\|\hat d\|\gamma _{g}\rangle$ is the one-electron dipole moment submatrix element. The 3\emph{d}-2\emph{p} hybrid structure of the even-parity molecular orbital:$\gamma_{g}\mu=N_{\gamma_g}(3d\gamma_{g}\mu+\lambda_{\gamma_g}2p\gamma_{g}\mu)$ and a  more simple form of purely oxygen odd-parity molecular orbital $\gamma_{u}\mu\equiv 2p\gamma_{u}\mu$ both with a symmetry superposition of the ligand O\,2\emph{p} orbitals point to a complex form of  the  submatrix element in  (\ref{d}) to be a sum of $local$ and $nonlocal$
terms composed of the one-site and two-site (\emph{d-p} and \emph{p-p}) integrals, respectively.
In the framework of a simple
"local" approximation\cite{Moskvin-Mn}, that implies the full neglect all many-center integrals:
$$
\langle t_{2u}(\pi )\|\hat d\|e_{g}\rangle =0;\, \langle t_{2u}(\pi )\|\hat
d\|t_{2g}\rangle = -i\sqrt{\frac{3}{2}}\lambda _{\pi}d \,;
$$
$$
\langle t_{1u}(\sigma )\|\hat d\|t_{2g}\rangle =0;\, \langle t_{1u}(\sigma
)\|\hat d\|e_{g}\rangle = -\frac{2}{\sqrt{3}}\lambda _{\sigma}d \, ;
$$
\begin{equation}
\langle t_{1u}(\pi )\|\hat d\|e_{g}\rangle =0;\, \langle t_{1u}(\pi )\|\hat
d\|t_{2g}\rangle = \sqrt{\frac{3}{2}}\lambda _{\pi}d\, .
\label{d-loc}
\end{equation}
Here, $\lambda _{\sigma}\sim t_{pd\sigma}/\Delta_{pd}$, $\lambda _{\pi}\sim t_{pd\pi}/\Delta_{pd}$ are $effective$ covalency parameters
 for $e_{g},t_{2g}$ electrons, respectively, $d=eR_0$ is an elementary dipole moment
 for the cation-anion bond length $R_0$.
 We see, that the "local" approximation results in an additional selection rule:
 it forbids the $\sigma \rightarrow \pi$, and  $\pi \rightarrow \sigma $
 transitions, $t_{1u}(\sigma )\rightarrow t_{2g}$, and
 $t_{1,2u}(\pi )\rightarrow e_{g}$, respectively, though these are dipole-allowed.
In other words, in frames of this approximation only $\sigma$-type
($t_{1u}(\sigma )\rightarrow e_{g}$) or $\pi$-type ($t_{1,2u}(\pi )\rightarrow
t_{2g}$) CT transitions are allowed.   Hereafter, we make use of the terminology of "strong" and
"weak" transitions for the dipole-allowed CT transitions going on the $\sigma
-\sigma$, $\pi -\pi$, and $\pi -\sigma$, $\sigma -\pi$ channels, respectively. It should be emphasized that the "local"
approximation, if non-zero, is believed to provide a leading contribution to transition
matrix elements with corrections being of the first order in the cation-anion
overlap integral. Moreover, the nonlocal terms are  neglected in  standard Hubbard-like approaches. 
In Fig.\,\ref{fig2} we do demonstrate the results of numerical calculations of several two-site dipole matrix elements against 3\emph{d} metal - oxygen separation ${\bf R}_{MeO}$.  It is clearly seen that given typical cation-anion separations ${\bf R}_{MeO}\approx 4 $ a.u. we arrive at values less than 0.1 a.u. even for the largest two-site integral, however, their neglect should be made carefully. Exps.(\ref{d}),(\ref{d-loc}) point to  likely extremely large dipole matrix elements and oscillator strengths for strong \emph{p-d} CT transitions, mounting to $d_{ij}\sim e\AA$ and $f\sim 0.1$, respectively. 

\subsection{One-center \emph{p-d} CT  transitions in manganites}

Now we can apply the model theory to the undoped stoichiometric perovskite manganites
RMnO$_3$. For our analysis to be more quantitative we make two rather obvious
model approximations. First of all, one assumes that for the MnO$_{6}^{9-}$ centers
in RMnO$_3$ as usually for cation-anion octahedra in  3\emph{d} oxides
\cite{Kahn,Licht,TMO} the non-bonding $t_{1g}(\pi)$ oxygen orbital has the
highest energy and forms the first electron removal oxygen state. Furthermore, to
be definite we assume that the energy spectrum of the non-bonding oxygen states
for Mn$^{3+}$O$_{6}^{9-}$ centers in RMnO$_3$
 coincides with that calculated in Ref.\onlinecite{Licht}
for Fe$^{3+}$O$_{6}^{9-}$ in isostructural orthoferrite LaFeO$_3$,
 in other words, we have (in eV):
 $$
 \Delta (t_{1g}(\pi)-t_{2u}(\pi))\approx 0.8\, ;\,\,
 \Delta (t_{1g}(\pi)-t_{1u}(\pi))\approx 1.8\, ;
 $$
 $$
 \Delta (t_{1g}(\pi)-t_{1u}(\sigma))\approx 3.0\, .
 $$
Secondly, we choose for the Racah parameters $B$ and $C$ the numerical values
typical for Mn$^{2+}$, $0.12$ and $0.41$ eV, respectively. The crystal-field parameter $Dq=0.15$ eV 
provides a reasonable explanation of the Mn$^{2+}$ spectra  in MnO \cite{MnO}. Furthermore, the photoemission data  \cite{Park} are believed to confirm the relevance of this value for crystal field splitting parameter in
RMnO$_3$. 
This set of parameters is used for the model theoretical simulation
of the overall \emph{p-d} CT band in LaMnO$_3$.\cite{Moskvin-Mn} Firstly, we argue that the lowest in
energy spectral feature observed in LaMnO$_3$
 near $1.7$ eV (see, e.g. Ref.\onlinecite{Jung}) is
believed to be related with the onset of  the series of the dipole-forbidden \emph{p-d}
CT transitions $t_{1g}(\pi)\rightarrow e_{g},t_{2g}$, rather than with any \emph{d-d}
crystal field transition. The energy of this transition was picked out to be a
starting point to assign all other \emph{p-d} CT transitions.

Weak dipole-allowed $\pi -\sigma$ CT transitions $t_{2u}(\pi)-e_{g}$ and
$t_{1u}(\pi)-e_{g}$ form more intensive CT bands starting at higher than the
preceding series energies,  near $2.5$  and $3.5$ eV, respectively, in
accordance with the magnitude of the $t_{1g}(\pi)-t_{2u}(\pi)$ and
$t_{1g}(\pi)-t_{1u}(\pi)$ separations. Actually, the $t_{1u}(\pi)-e_{g}$ 
transition at 3.5 eV has to be more intensive because the $t_{1u}(\pi)$ state is partly
hybridized with the $t_{1u}(\sigma)$ one, hence this transition borrows a portion of
intensity from the strong dipole-allowed $t_{1u}(\sigma )-e_g$ CT transition.

 The latter $\sigma -\sigma$ transition forms intensive broad \emph{p-d} CT $t_{2g}^{3}e_{g}^1\rightarrow t_{2g}^{3}e_{g}^2\underline{t}_{1u}(\sigma)$ band starting with the  main
${}^{5}E_{g}- {}^{6}A_{1g};{}^{5}T_{1u}$ peak at $\approx 4.7$ eV and ranging
to the  ${}^{5}E_{g}- {}^{4}A_{2g};{}^{5}T_{2u}$
 peak at $\approx 10.2$ eV with interstitial peaks at $\approx 8.0$ eV being the
 result of the superposition of two transitions
 ${}^{5}E_{g}- {}^{4}A_{1g};{}^{5}T_{1u}$ and
 ${}^{5}E_{g}- {}^{4}E_{g};{}^{5}T_{u}$, and at $\approx 8.8$ eV
 due to another ${}^{5}E_{g}- {}^{4}E_{g};{}^{5}T_{u}$ transition,
 respectively.
 Thus, the overall width of the \emph{p-d} CT bands with final
 $t_{2g}^{3}e_{g}^2$  configuration occupies a spectral range from
 $1.7$  up to $\sim 10$ eV.

 Strong dipole-allowed $\pi -\pi$ CT transitions
  $t_{2u}(\pi),t_{1u}(\pi)-
 t_{2g}$ form two manifolds of equally intensive $t_{2g}^{3}e_{g}^1\rightarrow t_{2g}^{4}e_{g}^1\underline{t}_{1,2u}(\pi)$ CT bands shifted with respect each other
 by the $t_{2u}(\pi)-t_{1u}(\pi)$ separation ($\approx 1.0$ eV).
 In turn, each manifold consists of two triplets  of weakly split and
 equally intensive \emph{p-d} CT bands
 related with ${}^{5}E_{g}- {}^{4}T_{1g};{}^{5}T_{u}$ and
 ${}^{5}E_{g}- {}^{4}T_{2g};{}^{5}T_{u}$ transitions, respectively.
 In accordance with the   assignment of crystal-field transitions
  \cite{MnO}  in LaMnO$_3$ (see Fig.\,1) we should expect the low-energy
  edge of the dipole-allowed $\pi -\pi$ CT band starting from $\approx 4.5$ eV
 ($1.7+2.0+$($t_{1g}(\pi)-t_{2u}(\pi)$ separation)). Taking account of strong
 configuration interaction we should expect the high-energy edge of this band
 related with the highest in energy ${}^{4}T_{2g}$ term of the 3\emph{d}$^5$
 configuration to be situated near $\approx 9.9$ eV. In between, in accordance
  with our scheme of energy levels we predict peaks at
 5.2, 5.5, 6.2 ($\times 2$),  7.2($\times 2$), 7.9, 8.2, 8.3, and 8.9 eV.
 The weak dipole-allowed $\sigma -\pi$ transitions occupy the high-energy
 spectral range from  $6.7$ to $11.1$ eV.

 Overall, our analysis shows  the multi-band structure of the \emph{p-d} CT optical response
 in LaMnO$_3$ with the weak low-energy edge at $1.7$ eV, related with the forbidden
$t_{1g}(\pi)-e_{g}$ transition and a series of  intensive bands in the range
$4.6\div 10.2$ eV starting with a composite peak at $\sim 4.5\div 4.7$ eV
and  closing by a composite peak at $8\div 10$ eV both resulting from
the superposition of strong dipole-allowed
$\pi -\pi$ and $\sigma -\sigma$ CT transitions.

In conclusion one should emphasize several points. First, the \emph{p-d} CT excitons can move over the lattice thus forming an excitonic band. Its width would strongly depend on the character of the O 2\emph{p} hole state. We anticipate rather narrow bands for $\pi -\pi$ and $\pi -\sigma$ excitons, but rather wide bands for $\sigma -\sigma$ and $\sigma -\pi$ excitons. 
It is worth noting that besides the charge transfer the \emph{p-d} CT transitions in MnO$_6$ octahedra are characterized by a remarkable effect of the  transfer of the spin density and orbital degeneracy from Mn ion to surrounding oxygen ions. In addition,  we would like to underline a crucial role of the intra-atomic correlations in forming the optical response in 3\emph{d} oxides. Indeed, the one-electron energy level scheme (see, Fig.\,1)  points to low-lying 3\emph{d}$t_{2g}$ orbital to be a final state for any lower-energy CT transition, while actually the lower-energy CT transitions are related with 3\emph{d}$e_{g}$ orbital as a final state because of a prevailing gain in the Hund energy. 

\subsection{Role of the Jahn-Teller distortions for the Mn$^{3+}$O$_6$ octahedra and the light polarization effects}

Orthorhombic  manganites RMnO$_3$ are  typical compounds with the 3\emph{d}-electron orbital ordered ground state where 3\emph{d}$_{3x^2-r^2}$-like and 3\emph{d}$_{3y^2-r^2}$-like
 orbitals for $e_g$-electrons are alternately ordered
on Mn$^{3+}$ sites in the $ab$ plane and are stacked parallel along
the c axis below the orbital ordering temperature T$_{JT}\sim$750-1500\,K\,\cite{Good2}. Such an orbital ordering is stabilized by the cooperative Jahn-Teller  (JT)  effect that lifts
the orbital degeneracy of the ground state crystal $^5$E term for the octahedral $t_{2g}^3e_g^2$ electronic configuration of Mn$^{3+}$ ion.
 The orbital ordering causes a layer-type  (so called A-type)  antiferromagnetic  AF  ordering below T$_N$, in which magnetic moments on Mn sites are
aligned ferromagnetically in the $ab$ plane and are stacked
antiferromagnetically along the $c$ axis.

The $e_g$-electron in the JT-distorted Mn$^{3+}$O$_6$ octahedral cluster occupies the superposition state
\begin{equation}
	\Psi _g=\cos\frac{\Theta}{2}|d_{z^2}\rangle + \sin\frac{\Theta}{2}|d_{x^2-y^2}\rangle \, ,
\end{equation}
where the orbital mixing angle  $\Theta$ is determined by the deformations of Mn$^{3+}$O$_6$ octahedral cluster as follows:
\begin{equation}
	\tan\Theta =\frac{\sqrt{3}(l_x-l_y)}{2l_z-l_x-l_y} \, ,
\end{equation}
where $l_{x,y,z}$ are the length of the Me-O bonds along the respective local co-ordinates.
The unoccupied, or $e_g$-hole state of the same Mn$^{3+}$O$_6$ octahedral cluster is described by the  wave function $\Psi_e$ which is orthogonal to $\Psi _g$:
\begin{equation}
\Psi _e=-\sin\frac{\Theta}{2}|d_{z^2}\rangle + \cos\frac{\Theta}{2}|d_{x^2-y^2}\rangle .
\end{equation}
The Jahn-Teller splitting energy is determined as follows: $\Delta_{JT}=2g\rho_0$, where $g$ is a vibronic constant,
\begin{equation}
\rho _0 =\left[2(l_x-l_y)^2+6(l_z-l)^2\right]^{1/2}\, ,
\label{ro}
\end{equation}
and $l$ is a mean Mn-O bond length.
For rare-earth orthorhombic  manganites the Mn-O$_I$ bond corresponds to the medium length while two Mn-O$_{II}$ bonds correspond to long and short lengths (see Ref.\onlinecite{Good2} and Fig.\,3). Interestingly that if the local $z$-axis is directed along Mn-O$_I$ bond and $x$-axis along  Mn-O$_{II}$ bond with the longest length, $(l_x-l)\gg (l_z-l)\approx (l_y-l)$ and $\tan\Theta\approx \sqrt{3}$, or $\Theta\approx 120^\circ$. In other words, it means that  $	\Psi _g \approx |d_{x^2}\rangle $. In the tetragonal limit ($(l_x-l)\gg (l_z-l)=(l_y-l)$) the $e_g$ level splits to two singlets ($|d_{x^2}\rangle$ and $|d_{y^2-z^2}\rangle$, respectively), while the $t_1,t_2$ levels split to  singlets and doublets (see Fig.\,1). Thus all the one-center \emph{p-d} CT transitions $t_{1,2}\rightarrow e_g$ from filled nonbonding oxygen orbitals to empty $e_g$ orbital, both allowed and forbidden, are believed to reveal a singlet-(quasi)doublet structure. The intensity of weakly dipole allowed $t_{2u}(\pi)\rightarrow e_g$  transitions centered near 2.5 eV,  can be  enhanced due to  a $t_{2u}(\pi)-t_{1u}(\pi),t_{1u}(\sigma)$ mixing resulting from the low-symmetry deformations of MnO$_6$ octahedra. 
\begin{figure}[t]
 \includegraphics[width=8.5cm,angle=0]{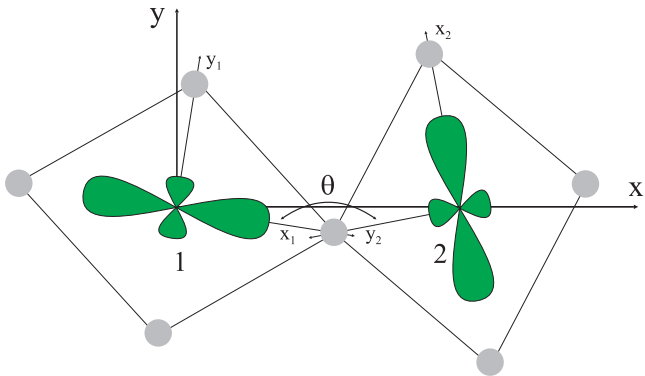}
 \caption{(Color online) Scheme of Mn$_1$-O$_{II}$-Mn$_2$ bonding in the $ab$-plane of perovskite manganites}
 \label{fig3}
 \end{figure} 
It should be noted that the probability for the $t_{1,2}\rightarrow t_{2g}$ transitions does not depend on the orbital mixing angle $\Theta$.

Let address the polarization properties of the main  dipole-allowed one-center \emph{p-d} CT transition $t_{1u}(\sigma)\rightarrow e_g$. To this end we find the dipole transition matrix elements in the local coordinates of MnO$_6$ octahedron:
\begin{widetext}
\begin{equation}
\langle t_{1ux}(\sigma)|d_x|d_{z^2}\rangle = \langle t_{1uy}(\sigma)|d_y|d_{z^2}\rangle =-\frac{1}{2}\langle t_{1uz}(\sigma)|d_z|d_{z^2}\rangle =-\frac{\lambda _{\sigma}d}{\sqrt{6}};\,\langle t_{1ux}(\sigma)|d_x|d_{x^2-y^2}\rangle =-\langle t_{1uy}(\sigma)|d_y|d_{x^2-y^2}\rangle =	\frac{\lambda _{\sigma}d}{\sqrt{2}}\, .
\end{equation}
and
\begin{equation}
\langle t_{1ux}(\sigma)|d_x|\Psi _e\rangle = \frac{\lambda _{\sigma}d}{\sqrt{6}}(\sin\frac{\Theta}{2}+\sqrt{3}\cos\frac{\Theta}{2});
\langle t_{1uy}(\sigma)|d_y|\Psi _e\rangle = \frac{\lambda _{\sigma}d}{\sqrt{6}}(\sin\frac{\Theta}{2}-\sqrt{3}\cos\frac{\Theta}{2});
\langle t_{1uz}(\sigma)|d_z|\Psi _e\rangle = -\frac{2\lambda _{\sigma}d}{\sqrt{6}}\sin\frac{\Theta}{2}\, .
\end{equation}
Hence for spectral weights we obtain the following relation:
\begin{equation}
	SW_x\,:SW_y\,:SW_z\,=(2+3\sin\Theta +\cos\Theta ):(2-3\sin\Theta +\cos\Theta ):2(1-\cos\Theta )\,.
\end{equation}
\end{widetext}
Neglecting the tilting of MnO$_6$ octahedra we arrive at the ratio of spectral weights SW$_c$ and SW$_{ab}$ for light polarized $\parallel$ $c$-axis and $\parallel$ $ab$-plane, respectively:
\begin{equation}
	SW_c/SW_{ab}=\frac{2(1-\cos\Theta )}{(2+\cos\Theta )}.
\end{equation}
This relation was obtained earlier by Tobe {\it et al.}\cite{Tobe} though the authors tried to apply it to explain the 2 eV band polarization properties. 
 Near the expected value of the orbital mixing angles ($\Theta \geq 108^{\circ}$) $SW_c/SW_{ab}\geq 1.5$, while the account for the tilting of MnO$_6$ octahedra is believed to reduce the ratio.

Generally speaking, the CT exciton creation is accompanied
by an excitation of lattice modes. Indeed, the electron transfer from O $2p$  to Mn
$3d$ state, or from an $ionic$ to a $covalent$
configuration such as $\sigma -\sigma$ transition ${}^{5}E_{g}- {}^{6}A_{1g};{}^{5}T_{1u}$ 
 is accompanied by a significant shortening of the
equilibrium Mn-O bond length and a remarkable effect of the  transfer of orbital degeneracy from Mn ion to surrounding oxygen ions accompanied by a strong change in the electron-lattice coupling.  Thus, the incident photon can create a self-trapped exciton which can reemit a photon, returning either to the
ground state or to various one-phonon or multiphonon excited states. The exciton-phonon interaction strongly affects the line-shape of absorption  leading to Franck-Condon multiphonon sidebands  and results in a phonon Raman scattering. The measurement of the Raman intensity as a function of excitation light energy is a very informative probe of the origin of electronic transitions. 

\subsection{Temperature dependence of the spectral weight for the \emph{p-d} CT bands}

Addressing the temperature dependence of the spectral weight for the \emph{p-d} CT transitions we will first concern the effects of a magnetic ordering. At first sight for small one-center excitons we have a rather conventional 
transition with conservation of spin $SM_S$-state and a  spin density fluctuation localized inside the MnO$_6$ cluster.  However, the redistribution of the spin density from manganese
atom to the oxygen ones after the \emph{p-d} CT transition switches on a strong ferromagnetic
Heisenberg O 2\emph{p}-Mn 3\emph{d} exchange that leads to a number of temperature anomalies near the N\'{e}el temperature. At first, one has to expect  a blue shift effect for the transition energy with the lowering the temperature near and below T$_N$. Indeed, at T$>$T$_N$, the average molecular field for the MnO$_6$ center turns into zero while the 3D antiferromagnetic ordering is accompanied by a rise of the exchange molecular fields and
respective spin splittings. Due to an order of magnitude bigger value of the O 2\emph{p}-Mn 3\emph{d} exchange as compared with the Mn 3\emph{d}-Mn 3\emph{d} exchange this is accompanied by an increase of the transition energy with a maximal
value of the blue shift as large as several
tenths of eV. Additionally, one has to expect a strong (of the same order of magnitude)
broadening of the excitonic line with the increase of the temperature due to strong
fluctuations of molecular fields. 
It is worth noting that at variance with the inter-site \emph{d-d} CT transition  the \emph{p-d} CT transition is not accompanied by strong two-magnon Raman processes, that could be used for its identification.

The temperature dependence of the phonon-assisted \emph{p-d} CT  transitions,  such as $t_{1g}(\pi)\rightarrow e_{g},t_{2g}$, normally forbidden by parity considerations, is usually described by the functional form predicted for such a
type of process:
$$
SW(T)=SW(0)\coth\frac{\hbar\omega_p}{2kT}\, ,
$$
where $SW(0)$, the spectral weight at 0 K, embodies the effect of parity mixing introduced by lattice vibrations in the states connected by the transition, and $\hbar\omega_p$ is a phonon energy. Depending on the energy of the active od\emph{d-p}arity phonon mode(s) such a mechanism can provide up to  a twofold rise of the spectral weight with the temperature rise from helium  to room temperatures. Obviously, this transition borrows a portion of intensity from the nearest in the energy dipole-allowed  CT transitions. 

It should be noted that equally with the band-structure effects the electron-phonon interaction governs the lineshape and the temperature dependence of all the CT transitions, both dipole-allowed and forbidden ones.  
 
\section{Two-center (\lowercase{\emph{d-d}}) CT transitions in manganites}

\subsection{Simple two-center CT model}

For illustration of covalence and correlation effects in a two-center MeO$_n$-MeO$_n$
molecular cluster, let first address its simplified version, namely a
two-electron two-center (A-B) system with a single on-site electron state.
Let a simple Hubbard-like Hamiltonian
\begin{widetext}
\begin{equation}
\hat H = t \sum _{\sigma}({\hat c}_{A\sigma}^\dagger{\hat
c}_{B\sigma}+ {\hat c}_{B\sigma}^\dagger{\hat c}_{A\sigma})+ U
\sum _{\sigma}(n_{A\sigma}n_{A-\sigma}+n_{B\sigma}n_{B-\sigma}) +V
\sum _{\sigma\sigma'}n_{A\sigma}n_{B\sigma'}
\end{equation}
\end{widetext}
incorporates a spin-independent single-particle
transfer ($t_{AB}=t_{BA}=t$) and the on-site, as well as inter-site
Coulomb repulsion. For a purely covalent bonding
($U=V=0;\, t\not= 0$) the conventional
single-particle MO-LCAO bonding-antibonding basis
set 
\begin{equation}
\Phi _{\pm}({\bf r}) = \frac{1}{\sqrt{2}}[\phi
_{A}({\bf r}) \pm \phi _{B}({\bf r})]
\label{MO}
\end{equation}
 provides a diagonalization of the transfer Hamiltonian, and we
have simple symmetrized
$$
\{\Phi _{+}\Phi _{+}\},\{\Phi _{+}\Phi _{-}\},\{\Phi _{-}\Phi
_{-}\}
 $$
two-particle basis set for spin singlets with energies $-|t|,0,+|t|$,
respectively, and antisymmetrized $[\Phi _{+}\Phi _{-}]$ function
for a spin triplet. All purely covalent states describe a bond-centered charge ordering  and are characterized by
the equal mean on-site electron density distribution, definite parity
and zero mean dipole moment. At the same time, there are strong
charge and dipole moment fluctuations.

Depending on the properties of the on-site functions  the
single-particle bonding $\Phi _{+}({\bf r})$ and antibonding
$\Phi _{-}({\bf r})$ MO-LCAO orbitals not only possess definite
parity with regard a center of inversion in the two-center cluster, but
resemble simple $s$- and $p$- orbitals, respectively. This implies
an occurrence of the specific for "covalent" systems mechanism of the
allowed electric-dipole transitions with the dipole matrix elements
determined only by the $AB$ separation
$$
\langle s|{\bf d}|p\rangle =\langle \Phi _{+}({\bf r})|{\bf d}|\Phi
_{-}({\bf r})\rangle = q{\bf R}_{AB}
$$
and observed in the ''longitudinal'' polarization ${\bf E} \parallel
{\bf R}_{AB}$.
\begin{figure}[t]
 \includegraphics[width=8.5cm,angle=0]{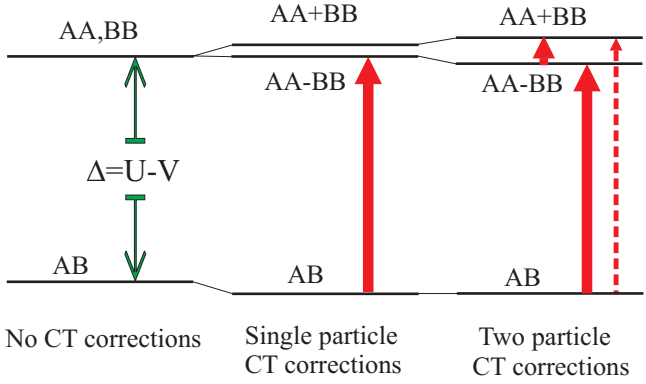}
 \caption{(Color online) Energy level scheme for simple two-center system with the step-by-step inclusion of single-particle and two-particle CT corrections. Bold arrow point to an allowed electro-dipole inter-site CT transition, dashed  arrow point to a forbidden  inter-site CT transition}
 \label{fig4}
 \end{figure}
In the opposite case of the purely ionic bonding ($U\not=V\not=0;\, t= 0$)
we apply a conventional Heitler-London approach forming two
one-site symmetrized configurations $\Psi _{AA}=\{\phi _{A}\phi
_{A}\}$ and $\Psi _{BB}=\{\phi _{B}\phi _{B}\}$ for spin singlets
which energy equals $U$, and two two-site configurations, described
by symmetrized even $\Psi _{AB}=\{\phi _{A}\phi _{B}\}$ and
antisymmetrized odd $\Psi _{[AB]}=[\phi _{A}\phi _{B}]$ functions
for spin singlet and triplet, respectively, which energies both
equal $V$. All initial purely ionic states describe a site-centered charge ordering with a strictly definite on-site electron density distribution.
Most interestingly that the $\Psi _{AA}$ and $\Psi _{BB}$ states interact due to a somewhat like an exchange resonance reaction $AA\leftrightarrow BB$ with a probability amplitude $T_{AB}=\langle \Psi
_{AA}|\hat H|\Psi _{BB}\rangle $, or  an effective  two-particle transfer integral. For a pair of identical centers we can compose even- and odd-parity superpositions  $\frac{1}{\sqrt{2}}(\Psi _{AA}\pm
\Psi _{BB})$  with $S$- and $P$-like behaviour, and the energy splitting due to the two-particle transfer ($\pm T_{AB}$), respectively.

Assuming $U>V$, and $|t|< \Delta =U-V$ we switch on a weak single-particle
inter-center transfer and simulate a weak covalent situation. In
frames of spin singlets we obtain three two-electron orbital states
(see Fig.\,4):

Even-parity ground state:
\begin{equation}
\Psi ^{g}_{GS}= \frac{1}{\sqrt{2}}\sin\theta
\,(\Psi _{AA}+\Psi _{BB})+ \cos\theta \,\Psi
_{AB};\,(E^{g}_{GS}\approx
V-\frac{t^{2}}{\Delta}); \label{GS1}
\end{equation}
Odd-parity excited state:
 \begin{equation}
\Psi ^{u}_{ES}= \frac{1}{\sqrt{2}}(\Psi
_{AA}-\Psi _{BB}); \quad (E^{u}_{ES}=U-T_{AB});
\label{odd}
\end{equation}
Even-parity excited state:
$$
\Psi ^{g}_{ES}= \frac{1}{\sqrt{2}}\cos\theta
\,(\Psi _{AA}+\Psi _{BB})- \sin\theta \,\Psi
_{AB};
$$
\begin{equation}
 (E^{g}_{ES}\approx
U+\frac{t^{2}}{\Delta}+T_{AB}), \label{even}
\end{equation}
where
\begin{equation}
\tan2\theta =\frac{2t}{\Delta}.
\label{tan}
\end{equation}
Here we ignore the spin-triplet state and ferromagnetic Heisenberg
exchange coupling.

Even-parity ground state $\Psi^{g}_{GS}$ and odd-parity excited state $\Psi^{u}_{ES}$ are coupled by a dipole-allowed charge transfer transition at the energy $\hbar \omega =\Delta+\frac{t^{2}}{\Delta}$ with a dipole matrix element
\begin{equation}
\langle \Psi ^{g}_{GS}|\hat{\bf d}|\Psi ^{u}_{ES}\rangle = 2e{\bf R}_{AB}\sin\theta	
\label{d_gu}
\end{equation}
We see that the oscillator strength or spectral weight for such a two-center \emph{d-d} CT transition
can be easily related  to the  kinetic contribution $J_{kin}=t^2/\Delta$ to the exchange integral in the AB
pair:$SW\propto J_{kin}/\Delta$\,\cite{Ahn,Oles}. 

Even- and odd-parity excited states $\Psi ^{g}_{GS}$ and  $\Psi ^{u}_{ES}$ are also coupled by a dipole-allowed  transition at the energy $\hbar\omega =\frac{t^{2}}{\Delta}+2|T_{AB}|$ with a dipole matrix element
\begin{equation}
\langle \Psi ^{g}_{ES}|\hat{\bf d}|\Psi ^{u}_{ES}\rangle = 2e{\bf R}_{AB}\cos\theta	
\end{equation}

The \emph{d-d} CT excitons can move over the lattice thus forming an excitonic band. Its width would strongly depend on the character of the electron-hole state. In particular, we anticipate rather wide bands for d$e_g$-d$e_g$ excitons, but rather narrow bands for d$t_{2g}$-d$t_{2g}$ excitons.


\subsection{Classification of two-center (\lowercase{\emph{d-d}}) CT transitions in manganites}
Simple model approach can be easily generalized to describe inter-center charge transfer transitions between two MeO$_n$
clusters centered at neighboring sites A and B, which  define two-center CT excitons in 3\emph{d} oxides. These two-center
excitons  may be addressed as quanta of the disproportionation
reaction
\begin{equation}
\mbox{MeO}_{n}^{v}+\mbox{MeO}_{n}^{v}\rightarrow
\mbox{MeO}_{n}^{v-1}+\mbox{MeO}_{n}^{v+1}\, ,
\label{r1}
\end{equation}
with the creation of electron  MeO$_{n}^{v-1}$ and hole
MeO$_{n}^{v+1}$ centers. Depending on the initial and final single particle  states all the intersite \emph{d-d} CT transitions  may be classified to the $e_g-e_g$, $e_g-t_{2g}$, $t_{2g}-e_g$, and $t_{2g}-t_{2g}$ ones.  For  the 3\emph{d} oxides with 3\emph{d} cations obeying the Hund rule these can be divided to so-called high-spin (HS) transitions $S_1S_2S\rightarrow S_{1}\pm \frac{1}{2}S_{2}\mp \frac{1}{2}S$ and low-spin (LS) transitions $S_1S_2S\rightarrow S_{1}- \frac{1}{2}S_{2}- \frac{1}{2}S$, respectively.

The inter-center \emph{d-d} CT transition in manganites 
\begin{equation}
\mbox{MnO}_{6}^{9-}+\mbox{MnO}_{6}^{9-}\rightarrow
\mbox{MnO}_{6}^{8-}+\mbox{MnO}_{6}^{10-}
\end{equation}
implies the  creation of the hole  MnO$_{6}^{8-}$ and  electron
MnO$_{6}^{10-}$ centers with electron configurations  formally related to Mn$^{4+}$ and Mn$^{2+}$, respectively. The HS one-particle \emph{d-d} CT transition driven by $e_g-e_g$ transfer can be written as follows:
$$
\mbox{MnO}_{6}^{9-}(t_{2g}^3;{}^4A_{2g}e_g^1;{}^5E_g)+\mbox{MnO}_{6}^{9-}(t_{2g}^3;{}^4A_{2g}e_g^1;{}^5E_g)\rightarrow
$$
\begin{equation}
\mbox{MnO}_{6}^{8-}(t_{2g}^3;{}^4A_{2g})+ \mbox{MnO}_{6}^{10-}(t_{2g}^3;{}^4A_{2g}e_g^2;{}^3A_{2g}:{}^6A_{1g}) \, .
\label{Mndd} 
\end{equation}
The transition energy $\Delta$ for such an anti-Jahn-Teller transition  is defined as follows: 
\begin{equation}
\Delta =\epsilon _{JT}+U_d-V_{dd}
\label{Delta}	
\end{equation}
that points to its dependence on the structural parameters (see (\ref{ro})).

The energies of different \emph{d-d} CT transitions $d^{n_1}d^{n_2}\rightarrow d^{n_1-1}d^{n_2+1}$ can be easily derived using the proper Tanabe-Sugano diagrams for the $d^{n_1-1}$ and $d^{n_2+1}$ configurations. 
For instance, the low-energy HS d($e_g$)-d($e_g$) CT transition ${}^5E_g{}^5E_g\rightarrow {}^4A_{2g}{}^6A_{1g}$ (\ref{Mndd}) sets up a rather wide band of the LS ${}^5E_g{}^5E_g\rightarrow {}^4A_{2g}{}^{4}\Gamma_2$ d($e_g$)-d($e_g$) CT transitions (${}^{4}\Gamma_2={}^{4}A_{1g},{}^{4}E_g({}^{4}G), {}^{4}E_g({}^{4}D), A_{2g}({}^{4}F)$) with the energy separations (given a cubic symmetry)\,\cite{Trees}:
$$
\Delta E({}^{4}A_{1g},{}^{4}E_g({}^{4}G))=10\,B+5\,C\,;
$$
$$
\Delta E({}^{4}E_g({}^{4}D))=17\,B+5\,C\,;
$$
$$
\Delta E({}^{4}A_{2g}({}^{4}F))=22\,B+7\,C\,\,,
$$ 
which do not depend on the crystal field splitting parameter $Dq$ and do scarcely vary from one compound to another that makes the transitions to be  important optical signatures of the d($e_g$)-d($e_g$) charge transfer. Making use for the Racah parameters  the numerical values typical for free Mn$^{2+}$ ion: B=0.12; C=0.41\,eV,  we arrive at three LS d($e_g$)-d($e_g$) CT bands separated from the low-energy HS d($e_g$)-d($e_g$) CT band by 3.2, 4.1, and 5.5 eV, while from experimental optical spectra for MnO\,\cite{MnO}
we obtain slightly lower energies: 2.8, 3.8, and 5.1 eV, respectively. 

It should be noted that the intra-atomic electron-electron repulsion does mix the $t_{2g}^{n_1};{}^{2S_1+1}\Gamma_1 e_g^{n_2};{}^{2S_2+1}\Gamma_2:{}^{2S+1}\Gamma$ states with different configurations and/or different intermediate momenta $S_{1,2},\Gamma_{1,2}$ but the same crystal terms ${}^{2S+1}\Gamma$. This point is of a great importance for ${}^{4}E_g({}^{4}G)$ and ${}^{4}E_g({}^{4}D)$ terms of Mn$^{2+}$ which are a result of the interaction of two bare terms: $t_{2g}^{3};{}^{4}A_{2g} e_g^{2};{}^{1}E:{}^{4}E_g$ and  $t_{2g}^{3};{}^{2}E e_g^{2};{}^{3}A_{2g}:{}^{4}E_g$, respectively. It is worth noting that only the former $t_{2g}^{3};{}^{4}A_{2g} e_g^{2};{}^{1}E:{}^{4}E_g$ term is active in the d($e_g$)-d($e_g$) charge transfer, which does not affect the $t_{2g}$ subshell.

The wave functions for ${}^{4}E_g({}^{4}G)$ and ${}^{4}E_g({}^{4}D)$ terms can be written as follows:
$$
\Psi ({}^{4}E_g({}^{4}D))=
$$
$$
\cos\alpha\,\Psi ({}^{4}A_{2g};{}^{1}E_g:{}^{4}E_g)+\sin\alpha\,\Psi ({}^{2}E_g;{}^{3}A_{2g}:{}^{4}E_g)	\\
$$
$$
\Psi ({}^{4}E_g({}^{4}G))=
$$
\begin{equation}
\sin\alpha\,\Psi ({}^{4}A_{2g};{}^{1}E_g:{}^{4}E_g)-\cos\alpha\,\Psi ({}^{2}E_g;{}^{3}A_{2g}:{}^{4}E_g)\,,	
\end{equation}
where $\tan 2\alpha =4\sqrt{3}$, and $\alpha \approx 41^{\circ}$. In other words, the both terms are almost equally involved in the d($e_g$)-d($e_g$) charge transfer under consideration.

In the $\approx 3$ eV gap in between the low-energy HS and LS d($e_g$)-d($e_g$) CT transitions one may observe  a relatively weak  low-energy HS d($t_{2g}$)-d($e_g$) CT transition ${}^5E_g{}^5E_g\rightarrow {}^4T_{2g}{}^6A_{1g}$ at the energy $\Delta +10Dq$ and low-energy LS d($e_g$)-d($t_{2g}$) CT transitions ${}^5E_g{}^5E_g\rightarrow {}^4A_{2g}{}^4T_{1g},{}^4T_{2g}$ anticipated at the energies $\Delta +2.1$\,eV and $\Delta +2.4$\,eV, respectively\,\cite{MnO}. Spectral weights for the both transitions are determined by 
$t_{2g}-e_g$($e_g-t_{2g}$) transfer integrals which are believed to be relatively small as compared with large $e_g-e_{g}$ transfer integrals.

\subsection{Two-center CT \emph{d-d} transitions:Effect of orbital states and Me$_1$-O-Me$_2$ bond geometry}
As it is seen from (\ref{tan}), (\ref{d_gu}) the \emph{d-d} CT transition probability amplitude  is determined by the respective single-particle transfer integral. These integrals depend both on initial and final states and the Me$_1$-O-Me$_2$ bond geometry. For two octahedral MeO$_6$ clusters sharing the common oxygen ion we arrive at following expressions for transfer integrals:\cite{thesis}
\begin{widetext}
$$
	t_{12}(e_g0;e_g0)\approx t_{ss}+t_{\sigma\sigma}\cos\theta;\,t_{12}(e_g0;e_g2)=t_{12}(e_g2;e_g0)=t_{12}(e_g2;e_g2)=0;\, 
$$
\begin{equation}	
t_{12}(e_g0;t_{2g}\mu)\approx t_{\sigma\pi}D_{0\mu}^{(1)}(\omega);\,t_{12}(t_{2g}\mu ; e_g0)\approx t_{\pi\sigma}D_{\mu 0}^{(1)}(\omega);\,t_{12}(t_{2g}\mu_1;t_{2g}\mu_2)\approx t_{\pi\pi}D_{\mu_1\mu_2}^{(1)}(\omega)\, ,
\end{equation}
\end{widetext}
where the 3\emph{d}-orbitals $e_g\mu $($e_g0=d_{z^2},e_g2=d_{x^2-y^2}$),$t_{2g}\mu$ ($t_{2g}\pm1 =\mp \frac{1}{\sqrt{2}}(d_{xz}\pm id_{yz}), t_{2g}2=d_{xy}$) are specified in the local co-ordinates for  Me$_1$O$_6$ and Me$_2$O$_6$ clusters with $z_1$ and $z_2$ axes directed to the common oxygen ion; $t_{ss,\sigma\sigma , \sigma \pi , \pi\sigma ,\pi\pi}$ are transfer parameters for $ss,\sigma\sigma , \sigma \pi , \pi\sigma ,\pi\pi$ bonds, respectively; $\theta$ is a Me$_1$-O-Me$_2$ bonding angle; $D_{\mu_1\mu_2}^{(1)}(\omega)$ is the Wigner rotation matrix\cite{Varshalovich} with $\omega$ being the Euler angles, specifying the transformation from the  local co-ordinates for  Me$_1$O$_6$ to that for Me$_2$O$_6$ cluster.

Thus we see that the  probability amplitude for the HS $\Psi _g(1)\rightarrow\Psi _e(2)$ CT $e_g\rightarrow e_g$ transition along $c$-axis in perovskite manganite RMnO$_3$ would depend on the crystallographic parameters as follows:
\begin{widetext}
\begin{equation}
A_c(\Psi _g(1)\rightarrow\Psi _e(2))=A_c(\Psi _g(2)\rightarrow\Psi _e(1))\propto \langle \Psi _g(1)|d_z|\Psi _e(2) \rangle \propto
\frac{t(\Psi _g(1)\rightarrow\Psi _e(2))}{\Delta}=\frac{1}{2}\sin\Theta \frac{(t_{ss}+t_{\sigma\sigma}\cos\theta _c)}{\Delta}
\end{equation}
\end{widetext}
while for the similar transition in the pair of nearest neighboring Mn$^{3+}$O$_6$ octahedral clusters located in $ab$-plane we arrive at
\begin{widetext}
 \begin{equation}
A_{ab}(\Psi _g(1)\rightarrow\Psi _e(2))\propto \langle \Psi _g(1)|d_{x,y}|\Psi _e(2) \rangle \propto
\frac{t(\Psi _g(1)\rightarrow\Psi _e(2))}{\Delta}=\frac{1}{2}(\frac{\sqrt{3}}{2}-\sin\Theta ) \frac{(t_{ss}+t_{\sigma\sigma}\cos\theta _{ab})}{\Delta};
\label{12}
\end{equation}
\begin{equation}
A_{ab}(\Psi _g(2)\rightarrow\Psi _e(1))\propto \langle \Psi _g(2)|d_{x,y}|\Psi _e(1) \rangle \propto
\frac{t(\Psi _g(2)\rightarrow\Psi _e(1))}{\Delta}=-\frac{1}{2}(\frac{\sqrt{3}}{2}+\sin\Theta ) \frac{(t_{ss}+t_{\sigma\sigma}\cos\theta _{ab})}{\Delta}\, .
\label{21}
\end{equation}
\end{widetext}
Hence  the spectral weight for the inter-site \emph{d-d} CT transitions may be written as follows:
$$
SW_c \propto \frac{1}{4}\sin^2\Theta \left(\frac{t_{ss}+t_{\sigma\sigma}\cos\theta _c}{\Delta}\right)^2;
$$
\begin{equation}
		SW_{ab}\propto \frac{1}{4}(\frac{3}{4}+\sin^2\Theta )\left(\frac{t_{ss}+t_{\sigma\sigma}\cos\theta _{ab}}{\Delta}\right)^2 \, .
\label{SW}
\end{equation}
It is worth noting that the orbital mixing angle dependence (\ref{SW})  strictly coincides with that of  Ref.\onlinecite{Kovaleva}.  Kim {\it et al.}
\cite{Kim2} presented an expression  for SW$_{ab}$ which accounts only for $1\rightarrow 2$ transfer (compare our Exp.(\ref{12}) and Exp.(4) from Ref.\onlinecite{Kim2}) that immediately questions their main conclusions.

The spectral weight for the two-center d$e_g$-d$e_g$ CT transition can be easily related  with the  kinetic contribution to the $e_g-e_g$ exchange integral in the Mn$^{3+}$-Mn$^{3+}$
pair:
$$
SW_c \propto \frac{J_{kin}^c(e_g-e_g)}{\Delta}\sin^2\Theta ;
$$
\begin{equation}
		SW_{ab}\propto \frac{J_{kin}^{ab}(e_g-e_g)}{\Delta}(\frac{3}{4}+\sin^2\Theta ) \, .
\label{SWabc}
\end{equation}
The both expressions do predict a sizeable suppression of the SW$_{ab}$  and SW$_{c}$ in RMnO$_3$ for the two-center d$e_g$-d$e_g$ CT transition with decreasing the ionic radius of R-ion, if to take into account proper variation of the structural parameters $\theta$, $\Theta$, and $\Delta _{JT}\propto \rho$\,\cite{Good2}.

\subsection{Two-center CT \emph{d-d} transitions:Spin dependence and temperature behaviour}
After ruling out the roles of the Mn-O-Mn bond angle and the octahedral-site distortion in determining the \emph{d-d} CT transition probability, we have to consider whether the spin degree of freedom affect the \emph{d-d} CT transition.

Interestingly that despite the spinless character of dipole moment operator its matrix on the pair wave functions depends on the spin quantum numbers
\begin{widetext}
\begin{equation}
\langle \Phi_1\Phi_2S_1S_2SM_S|{\hat d}_{q}|\Phi_1^{\prime}\Phi_2^{\prime}S_{1}^{\prime}S_{2}^{\prime}S^{\prime}M_{S^{\prime}}\rangle = (-1)^{S}\delta _{SS^{\prime}}\delta _{M_{S}M_{S^{\prime}}}\sqrt{[S_1,S_{2}^{\prime}]}
\left\{ \begin{array}{ccc}
S_1 & S_2 & S \\
S_{2}^{\prime} & S_{1}^{\prime} & \frac{1}{2}
\end{array}
\right\} \langle \Phi_1\Phi_2|{\hat d}_{q}|\Phi_1^{\prime}\Phi_2^{\prime}\rangle\, ,
\end{equation}
\end{widetext}
for the $1\rightarrow 2$ ($S_1\rightarrow S_{2}^{\prime}$) electron transfer. Here the conventional notations $\left\{ \begin{array}{ccc} \cdot & \cdot &
\cdot \\ \cdot & \cdot & \cdot
\end{array}
\right\} $ are used for spin $6j$-symbol, $[S]=2S+1$. For the partial spectral weight for $S_1S_2S\rightarrow S_{1}^{\prime}S_{2}^{\prime}S$ transition we have
\begin{widetext}
\begin{equation}
SW(S_1S_2S\rightarrow S_{1}^{\prime}S_{2}^{\prime}S)\propto [S]^{-1}\sum_{M_S}|\langle \Phi_1\Phi_2S_1S_2SM_S|{\hat d}_{q}|\Phi_1^{\prime}\Phi_2^{\prime}S_{1}^{\prime}S_{2}^{\prime}SM_{S}\rangle|^2 = [S_1,S_{2}^{\prime}]
\left\{ \begin{array}{ccc}
S_1 & S_2 & S \\
S_{2}^{\prime} & S_{1}^{\prime} & \frac{1}{2}
\end{array}
\right\}^2 |\langle \Phi_1\Phi_2|{\hat d}_{q}|\Phi_1^{\prime}\Phi_2^{\prime}\rangle|^2\, ,
\end{equation}
where
\begin{equation}
\left\{ \begin{array}{ccc}
S_1 & S_2 & S \\
S_{2}+ \frac{1}{2}& S_{1}- \frac{1}{2} & \frac{1}{2}
\end{array}
\right\} \,= (-1)^{S_1+S_2+S}\left[\frac{S(S+1)-(S_1-S_2)(S_1-S_2-1)}{2S_1(2S_1+1)(2S_2+1)(2S_2+2)}\right]^{1/2},
\end{equation}
\begin{equation}
\left\{ \begin{array}{ccc}
S_1 & S_2 & S \\
S_{2}- \frac{1}{2}& S_{1}- \frac{1}{2} & \frac{1}{2}
\end{array}
\right\} \,= (-1)^{S_1+S_2+S}\left[\frac{(S_1+S_2)(S_1+S_2+1)-S(S+1)}{2S_1(2S_1+1)2S_2(2S_2+1)}\right]^{1/2},
\end{equation}
\end{widetext}

\begin{figure}[t]
 \includegraphics[width=8.5cm,angle=0]{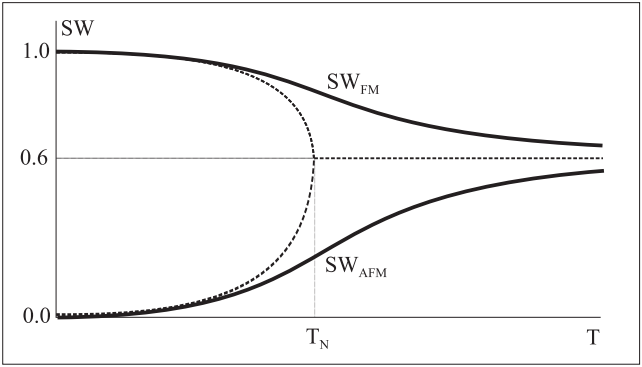}
 \caption{Schematic temperature dependence of the spectral weight for the HS \emph{d-d} CT transition in ferromagnetically and antiferromagnetically ordering pairs, respectively}
 \label{fig5}
 \end{figure}

Making use of the normalization relation for $6j$-symbols\cite{Varshalovich}
\begin{equation}
\sum_{S_{2}^{\prime}}[S_1,S_{2}^{\prime}]\left\{ \begin{array}{ccc}
S_1 & S_2 & S \\
S_{2}^{\prime} & S_{1}^{\prime} & \frac{1}{2}
\end{array}
\right\}^2=	1
\end{equation}
\begin{widetext}
we arrive at a spin sum rule:
\begin{equation}
 [S]^{-1}\sum_{S_{2}^{\prime}M_S}|\langle \Phi_1\Phi_2S_1S_2SM_S|{\hat d}_{q}|\Phi_1^{\prime}\Phi_2^{\prime}S_{1}^{\prime}S_{2}^{\prime}SM_{S}\rangle|^2 =  |\langle \Phi_1\Phi_2|{\hat d}_{q}|\Phi_1^{\prime}\Phi_2^{\prime}\rangle|^2\, ,
\label{sr}
\end{equation}
For the partial spectral weight for $S_1S_2\rightarrow S_{1}^{\prime}S_{2}^{\prime}$ transition we have 
\begin{equation}
SW(S_1S_2\rightarrow S_{1}^{\prime}S_{2}^{\prime})\propto [S_1,S_{2}^{\prime}]\sum_S \rho _S
\left\{ \begin{array}{ccc}
S_1 & S_2 & S \\
S_{2}^{\prime} & S_{1}^{\prime} & \frac{1}{2}
\end{array}
\right\}^2 |\langle \Phi_1\Phi_2|{\hat d}_{q}|\Phi_1^{\prime}\Phi_2^{\prime}\rangle|^2\, ,
\end{equation}
\end{widetext}
where $\rho _S$ is the temperature dependent statistical weight of $S_1S_2S$ spin multiplet.
Taking into account the above expressions for $6j$-symbols we see that the temperature dependence of the partial spectral weight $SW(S_1S_2\rightarrow S_{1}^{\prime}S_{2}^{\prime})$ would be determined by a statistical average $\langle S(S+1)\rangle=\langle \hat {\bf S}^2\rangle$ which, in its turn, relates to  the spin-spin correlation function $\left\langle\left(\hat {\bf S}_1\cdot \hat {\bf S}_2\right)\right\rangle$ as follows
\begin{equation}
	\left\langle\left(\hat {\bf S}_1\cdot \hat {\bf S}_2\right)\right\rangle=\frac{1}{2}\left[\langle S(S+1)\rangle-S_1(S_1+1)-S_2(S_2+1)\right]
\end{equation}
Thus we should conclude that the partial spectral weight for $S_1S_2S\rightarrow S_{1}^{\prime}S_{2}^{\prime}S$ transition in an isolated spin pair is governed by a spin-dependent prefactor containing the spin-spin correlation function:$\left\langle\left(\hat {\bf S}_1\cdot \hat {\bf S}_2\right)\right\rangle$. For the HS transition $S_1S_1\rightarrow S_1-\frac{1}{2}S_1+\frac{1}{2}$ and the LS transition $S_1S_1\rightarrow S_1-\frac{1}{2}S_1-\frac{1}{2}$ in the pair of identical 3\emph{d} ions we arrive at
\begin{widetext}
\begin{equation}
	SW(S_1S_1\rightarrow S_1-\frac{1}{2}S_1+\frac{1}{2})\propto \frac{\langle S(S+1)\rangle}{2S_1(2S_1+1)}=\frac{\left[\left\langle\left(\hat {\bf S}_1\cdot \hat {\bf S}_2\right)\right\rangle+S_1(S_1+1)\right]}{S_1(2S_1+1)} \, ,
\label{SWHS}
\end{equation}
\begin{equation}
	SW(S_1S_1\rightarrow S_1-\frac{1}{2}S_1-\frac{1}{2})\propto \frac{\left[2S_1(2S_1+1)-\langle S(S+1)\rangle \right]}{2S_1(2S_1+1)}=\frac{\left[S_1^2-\left\langle\left(\hat {\bf S}_1\cdot \hat {\bf S}_2\right)\right\rangle\right]}{S_1(2S_1+1)} \, ,
\label{SWLS}
\end{equation}
\end{widetext}
respectively. These expressions allow to obtain easily both the low-temperature ($T\ll T_N$) and high-temperature ($T\gg T_N$, $\left(\hat {\bf S}_i\cdot \hat {\bf S}_j\right)\rightarrow 0$) limits for the spin prefactor. Schematically the temperature dependence of the spectral weight for the HS \emph{d-d} CT transition in ferromagnetically and antiferromagnetically ordering Mn$^{3+}$-Mn$^{3+}$ pairs ($S_1=2$) is shown in Fig.\ref{fig5}. It should be noted that similar results were obtained earlier by Ole\'s {\it et al.}\,\cite{Oles} though from another starting point.

In accordance with a spin sum rule (\ref{sr}) the sum of spin prefactors in the right hand sides of the expressions (\ref{SWHS}) and (\ref{SWLS}) turns into unity due to an exact compensation of temperature dependent terms with the spin-spin correlation function. In other words, the higher-energy LS bands exhibit a strictly opposite SW evolution with the temperature as compared with HS band, that is varying the temperature we arrive at the SW transfer between the HS- and LS-subbands. In the limit $T\rightarrow 0$ the HS band in the antiferromagnetically coupled spin pair or LS band in the ferromagnetically coupled spin pair  are freezed out. However, this prohibition is lifted for the HS band in the antiferromagnetically coupled spin pair, if we take account of effective staggered magnetic fields which affect the spin pair in the antiferromagnetic surroundings giving rise to a spin singlet-triplet mixing. Furthermore, the low-temperature SW for  the HS band in the antiferromagnet is believed to be a quantitative measure of the N\'eel state weight in the ground antiferromagnetic state. 
 The ratio of spectral weights $SW_c$ and $SW_{ab}$ does weakly depend on the structural parameters such as $\Theta$ and  $\theta$, in the high-temperature limit ($T\gg T_N$) $SW_c/SW_{ab}\approx 0.5$, while in the low-temperature limit ($T\ll T_N$) for antiferromagnetic A-type order this ratio turns into zero.
  It is important to note that the temperature variation between these two limits
is controlled by the nearest-neighbor spin-spin correlation function, which varies smoothly with T, and not by the square of the sublattice magnetization, which of course vanishes above T$_N$.
One generally has considerable short-range order above any magnetic-ordering temperature, and for LaMnO$_3$ with $T_N=140$\,K the ferromagnetic correlations should persist to quite high temperatures and the full difference in the spectral weight of the \emph{d-d} CT transition between the antiferromagnetic ground state and the paramagnetic state is not saturated by 300\,K. In contrast with LaMnO$_3$ the room temperature seems to correspond quite well to a high-temperature limit ($T\gg T_N$) for such  manganites as SmMnO$_3$ and EuMnO$_3$ with $T_N\approx 60$ and 50\,K, respectively\cite{Good2}. In other words, the  optical measurements could be used to inspect the short-range spin correlations in manganites.
It is worth noting that the spin-spin correlation function  $\left\langle\left(\hat {\bf S}_i\cdot \hat {\bf S}_j\right)\right\rangle$   for the nearest-neighbor spins localized on the i-th and j-th Mn sites governs the effect of phonon renormalization\,\cite{Granado}. This effect  was examined \cite{Granado,Mukhin} for all orthorhombic RMnO$_3$ compounds.

\subsection{Some features of CT transitions in strongly coupled corner-shared MnO$_6$ clusters}

Above we have addressed a somewhat idealized scenario of \emph{p-d} and \emph{d-d} CT transitions which implies well isolated or weakly coupled MnO$_6$ clusters. Actually in perovskite manganites we deal with strongly coupled corner-shared MnO$_6$ clusters sharing a common oxygen ion. Strictly speaking, it means that we cannot make use of an oversimplified classification of the \emph{p-d} and \emph{d-d} CT transitions. Indeed, the O 2\emph{p} electrons localized on the intermediate common oxygen ion cannot be attributed to one or another cluster. Formally this implies a strong overlap of the bare wave functions localized on the neighboring clusters. Generally speaking, this trouble can be overcomed with making use of the orthogonalization procedure, e.g. by the L\"{o}wdin technique. This immediately results in a two-center character of the wave functions with the onset of two-center spin and orbital correlations. 

Many features of the corner-shared MeO$_6$ clusters related with the \emph{p-d} overlap and covalency effects can be illustrated by a more simple generic two-hole-three-site  M$_1$-O-M$_2$ cluster model to be a natural generalization of the above addressed two-site A-B model. 
 To this end we make use of a technique suggested in Refs.\cite{DM-JETP,DM-PRB,mechanism} to derive the expressions for the Dzyaloshinsky-Moriya coupling and spin-dependent electric polarization in copper oxides. 

We start with the construction of spin-singlet and spin-triplet wave functions for our three-center two-hole system taking account of the \emph{p-d} hopping, on-site hole-hole repulsion, and crystal field effects for the bare ground state 101 and bare  excited configurations $\{n\}$ (011, 110, 020, 200, 002) with different hole occupation of M$_1$, O, and M$_2$ sites, respectively. The \emph{p-d} hopping for M-O bond implies a conventional Hamiltonian
 \begin{equation}
 \hat H_{pd}=\sum_{\alpha \beta}t_{p\alpha d\beta}{\hat p}^{\dagger}_{\alpha}{\hat d}_{\beta}+h.c.\, ,
 \label{Hpd}
\end{equation}
 where ${\hat p}^{\dagger}_{\alpha }$ creates a hole in the $\alpha $ state on the oxygen site, while  ${\hat d}_{\beta}$
annihilates a hole  in the $\beta $ state on the copper site.
 
 Perturbed wave functions can be written as follows
\begin{equation}
\Psi_{\{n\}\Gamma SM}=\eta^{\Gamma S}_{\{n\}}[\Phi_{\{n\}\Gamma SM}+\sum_{\{n\}^{\prime}\not=\{n\},\Gamma^{\prime}}c_{\{n\}\Gamma S}^{\{n\}^{\prime}\Gamma^{\prime} S}\Phi_{\{n\}^{\prime}\Gamma^{\prime} SM}],
	\label{Psi_1}
\end{equation}
where $\Phi_{\{n\}\Gamma SM}$ are bare wave functions for $\{n\}$ configuration and the summation runs both on different configurations and different orbital $\Gamma$ states;
\beq
\eta^{\Gamma S}_{\{n\}}=\left(1+\sum_{\{n\}^{\prime}\not=\{n\},\Gamma^{\prime}}\left|c_{\{n\}\Gamma S}^{\{n\}^{\prime}\Gamma^{\prime}S}\right|^2\right)^{-1/2}
\label{norm}
\eeq
is a normalization factor. If we neglect the overlap integrals, the probability amplitudes for configurations coupled by a single \emph{p-d} transfer are defined by the ratio of the generalized \emph{p-d} transfer integrals and the \emph{p-d} transfer energy as follows:
\begin{equation}
c_{\{n\}\Gamma S}^{\{n\}^{\prime}\Gamma^{\prime} S}=-\frac{\langle \{n\}^{\prime}\Gamma^{\prime} S|\hat H|\{n\}\Gamma S\rangle}{\Delta (\{n\}^{\prime}\Gamma^{\prime} S-\{n\}\Gamma S)}	\, .
\end{equation}
In other words, $|c_{101}^{011}|\sim |c_{101}^{110}|\sim |c_{020}^{110}|\sim |c_{110}^{020}|\sim |c_{110}^{200}|,...\sim |\frac{t_{pd}}{\Delta_{pd}}|$, while the probability amplitudes, or hybridization parameters for configurations coupled by a two-step \emph{p-d} transfer, such as  $c_{101}^{200}, c_{101}^{020}, c_{101}^{002}$ are on the order of $|\frac{t_{pd}}{\Delta_{pd}}|^2$.

Formally, the 101$\rightarrow$011(110) and 101$\rightarrow$002(200) CT transitions may be attributed to \emph{p-d}  and \emph{d-d} charge transfer, respectively. However, the configuration interaction due to the charge transfer and overlap do result in that both types of the CT transitions are accompanied by a charge redistribution all over the three centers. Furthermore, a novel type of \emph{p$^2$-dd} CT transitions 101$\rightarrow$020 does emerge due to a mixing of all the 101, 110, 011, 020 configurations in the initial and final states. Obviously, the main contributions to dipole transition matrix elements for \emph{p-d}, \emph{d-d}, and \emph{p$^2$-dd} CT transitions are proportional to  $\frac{t_{pd}}{\Delta_{pd}}$, $|\frac{t_{pd}}{\Delta_{pd}}|^2$, and $|\frac{t_{pd}}{\Delta_{pd}}|^2$, respectively. Accordingly, the spectral weights for the dipole allowed \emph{p-d}, \emph{d-d}, and \emph{p$^2$-dd} CT transitions are 
$$
SW_{p-d}\propto \left(\frac{t_{pd}}{\Delta_{pd}}\right)^2;\, SW_{d-d}\propto  SW_{p^2-dd}\propto \left(\frac{t_{pd}}{\Delta_{pd}}\right)^4,
$$
 respectively. 

 Interestingly, all the dipole-allowed CT transitions obey the $\Delta S=0$ spin-selection rule, however, the dipole transition matrix elements depend on the total spin S  due to a spin (S) dependence of the energy denominators in the probability amplitudes and so in the normalization factors, particularly due to a contribution of 200, 020, and 002 two-hole on-site configurations. Indeed, for such configurations the spin-singlet and spin-triplet terms are usually separated by a large energy gap. In general, the  spin-dependent part of the electro-dipole moment operator acting in the basis of the three-site wave functions   can be written as follows      
\begin{equation}
	\hat{\bf P}_s=\hat{\bf \Pi}(\hat{\bf s}_1\cdot \hat{\bf s}_2)\,.
	\label{TMS}
\end{equation}
Such a spin operator for a transition dipole moment was firstly introduced by Sugano {\it et al}.\,\cite{TMS} to explain so-called magnon side-bands in the optical absorption spectra of magnetic crystals. Simple estimates yield $|{\bf \Pi}|\propto \left(\frac{t_{pd}}{\Delta_{pd}}\right)^3$ and $|{\bf \Pi}|\propto \left(\frac{t_{pd}}{\Delta_{pd}}\right)^2$ for \emph{p-d} and \emph{d-d} CT transtions, respectively. In other words, the spin-dependent part of the spectral weight for both transitions appears to be of the same order: $SW_{spin}\propto \left(\frac{t_{pd}}{\Delta_{pd}}\right)^4$. It should be noted that the spin-dependent part of the spectral weight in the three-site two-hole cluster can be written as follows:   
\begin{equation}
	SW_{spin}=(SW_{triplet}-SW_{singlet})(\hat{\bf s}_1\cdot \hat{\bf s}_2)\,,
	\label{TMSa}
\end{equation}
if to take into account that 
$$
(\hat{\bf s}_1\cdot \hat{\bf s}_2)^2=\frac{3}{16}-\frac{1}{2}(\hat{\bf s}_1\cdot \hat{\bf s}_2). 
$$
 From this reasoning we may conclude that, in contrast to  a clue point of papers by Kovaleva {\it et al}.\,\cite{Kovaleva,Kovaleva-09},  the both dipole-allowed \emph{d-d} and \emph{p-d} CT transtions are equally sensitive to the temperature-dependent spin correlations, only with a different $relative$ change of the optical response through the onset of the magnetic order, {\it relatively large} for the former and {\it relatively small} for the latter.

\section{Ellipsometric study of optical response of RM\lowercase{n}O$_3$}
Theoretical analysis (see preceding section and Refs.\onlinecite{Kovaleva,Kovaleva-09}) clearly shows that the spectral weight of the HS d$e_g$-d$e_g$ CT transitions is to be sizeably suppressed with decreasing the ionic radius of R-ion and increasing structural distortions that provides an opportunity to a better manifestation and inspection of the relatively weak \emph{p-d} CT transitions expectedly forming a fine structure of 2 eV band in manganites.    Indeed, a marked suppression of the spectral weight for the 2 eV band was observed by Kim {\it et al}.\,\cite{Kim2}  in the ab-plane absorption spectra of RMnO$_3$ (R=La, Pr, Nd, Gd, and Tb) thin films. However, the authors focused on the suppression effect and did not concern the fine structure of the band.   
 We have, on the contrary, studied the optical response of RMnO$_3$ (R=La, Pr, Nd, Sm, and Eu) with a main interest in the fine structure of 2 eV band and a more comprehensive assignment of different \emph{p-d} and \emph{d-d} CT transitions.

The  perovskite manganites RMnO$_3$ crystallize in the orthorhombic structure  (space group $Pbmn$), where Mn$^{3+}$O$_6$ octahedra share corners with Mn$^{3+}$-O-Mn$^{3+}$ angle varying from  $\approx 155^{\circ}$ for LaMnO$_3$ to $\approx 140^{\circ}$ in the end of the series\,\cite{Good2}.
The orthorhombic RMnO$_3$ compounds with large ionic radia from La to Gd  are characterized by A-type antiferromagnetic AFM  structure. In this structure the Mn moments are ferromagnetically  (FM)  ordered in the $ab$  planes and antiferromagnetically coupled along the $c$  direction. The N\'eel temperature T$_N$ decreases from 140 K for LaMnO$_3$ to
40 K for GdMnO$_3$\,\cite{Good2}.
 With increasing structural distortions, the magnetic structure of RMnO$_3$
compounds with small ionic radii   changes to an incommensurate  IC  sinusoidal AFM structure (R=Tb, Dy) or to a so-called E-structure (R=Ho, Er). 

We have studied the optical response for single crystalline samples of RMnO$_3$ (R=La, Pr, Nd, Sm, Eu)  grown by the floating-zone technique described in Ref.\onlinecite{Balbashov}. Magnetic properties of these samples are reported in Refs.\onlinecite{Mukhin1,Kadomtseva}.
As a rule, the as-grown single crystals of RMnO$_3$ had twins, however, it was possible to choose 
some samples exhibiting a strong magnetic and optical anisotropy, hence these could be considered as practically untwinned.

\begin{figure*}[t]
 \includegraphics[width=17.0cm,angle=0]{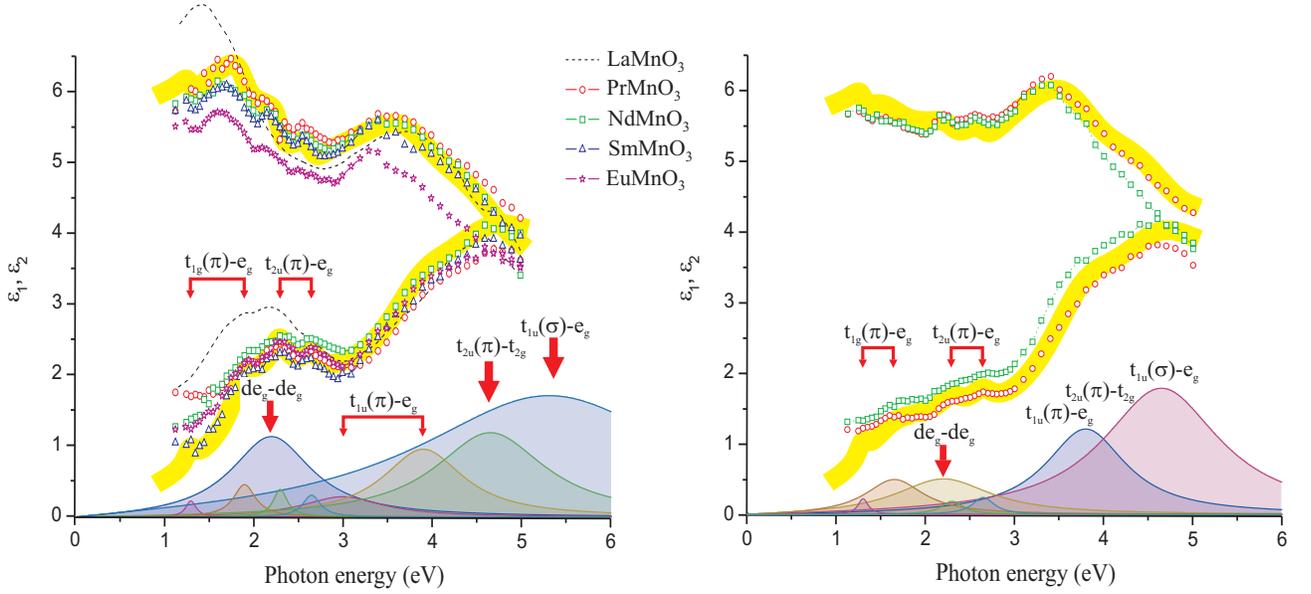}
 \caption{(Color online) Real and imaginary part of the dielectric function $\varepsilon $ in RMnO$_3$. The left hand panel presents the spectra for $\bf E\perp \bf c$ polarization, the right hand panel does for $\bf E\parallel \bf c$ polarization. The Lorentzian fitting data are shown by solid lines. A proper Lorentzian decomposition with assignment of all the peaks is shown for $\varepsilon_2 $. }
 \label{fig6}
 \end{figure*} 
 The measurements were performed  with a home-built automatic ellipsometer on the x-ray oriented single crystalline bulk samples with as-grown polished surface ($\sim 3\times 5$ mm$^2$) where the angle of incidence was 67$^{\circ}$ or 72$^{\circ}$. Spectroscopic ellipsometry measurements were performed on the $àñ$ surfaces for PrMnO$_3$ and NdMnO$_3$ crystals for high-symmetry orientations $a$ and $c$ of the optical axis with respect to the plane of incidence of light and on the $ab$ surfaces for SmMnO$_3$ and EuMnO$_3$ crystals for high-symmetry orientation $a$.  The crystal surfaces ($ac$-plane for PrMnO$_3$ and NdMnO$_3$, and $ab$-plane for SmMnO$_3$ and EuMnO$_3$) were polished to optical grade with diamond powders.

Optical complex dielectric function $\varepsilon=\varepsilon_{1}-i\varepsilon_{2}$ of single crystalline samples of RMnO$_3$ was studied at room temperature in the spectral range from 0.5 to 5.5\, eV. However, our experimental technique had some spectral limitations and provided reliable optical data only in the spectral range 1.5-5 eV of strong absorption. 
 Optical constants, refractive index $n$ and extinction ratio $k$ have been calculated with using relations reported in Ref.\onlinecite{Nomer}. 
 The technique of ellipsometry provides significant advantages over conventional reflection methods in that it is self-normalizing and does not require reference measurements, and optical complex dielectric function
$\varepsilon=\varepsilon_{1}-i\varepsilon_{2}$  are obtained directly
without a Kramers-Kr\"onig transformation. The comparative analysis of the spectral behavior of $\varepsilon_{1}$ and $\varepsilon_{2}$ is believed to provide a more reliable assignement of spectral features.


Fig.\,\ref{fig6}  shows the $\varepsilon_{1}$ and $\varepsilon_{2}$
room temperature spectra of RMnO$_3$ (R= Nd, Pr, Sm, Eu) for two  main light polarizations, $E\perp {\bf c}$-axis and $E\parallel {\bf c}$-axis, respectively, together with earlier data for LaMnO$_3$\cite{optics-PS}. 
The optical spectra of RMnO$_3$ are interesting in several aspects. 
First of all, as in many earlier works on LaMnO$_3$, we clearly see  two intensive and rather broad optical features peaked at around 2 eV and 4.5 eV, respectively. Both the intensity and shape of the high-energy band change weakly inbetween different manganites, whereas the spectral weight of the broad band peaked at 2.0 eV in  LaMnO$_3$ seems by  factor 1.5 bigger than in other manganites. This result  agrees with the findings by Kim {\it et al.}\cite{Kim2} who investigated the $ab$-plane absorption spectra of RMnO$_3$ (R=La,Pr,Nd,Gd, and Tb) thin films and observed a drastic suppression of the 2 eV peak with the decrease of the R-ion radii.


The broad 2 eV band exhibits strong anisotropy, it is hardly visible at $\bf E\parallel \bf c$-axis polarization. 
As expected, our measurements for R=Nd, Pr, Sm, Eu have  uncovered many subtle features missed in earlier optical experiments on parent manganites. First, the comparative analysis of the $\varepsilon_{1}$ and $\varepsilon_{2}$ spectra (see Fig.\,\ref{fig6}) allows us to unambiguously conclude that the 2 eV feature is composed of a single intensive and rather broad band peaked at 2.0 eV for   LaMnO$_3$ and 2.2 eV for other manganites, and several relatively weak and narrow bands peaked nearly equally for all the manganites at 1.3, 1.9, 2.3, 2.7 eV ($\varepsilon_{ab}$), respectively. Interestingly, that not only the energies but also the intensities of these bands hardly depend on the manganite type if any. This observation immediately points to different nature of both features  at variance with  Kovaleva {\it et al.}\cite{Kovaleva} who addressed the 2 eV feature in LaMnO$_3$ to consist of three distinct bands, peaked at 1.95, 2.35, 2.70  ($\varepsilon _{2ab}$, T=300\,K) each attributed to inter-site \emph{d-d} CT transitions. 
 Weak sub-peaks show up a clear anisotropy.

As regards the high energy spectral feature peaked at around 4.5 eV one should note its hardly visible composite structure with as minimum three rather intensive bands which integral spectral weight for $\bf E\parallel \bf c$-axis polarization seems to be  slightly larger than for $\bf E\perp \bf c$-axis polarization.


Our ellipsometric data for RMnO$_3$ samples (R=Pr, Nd, Sm, Eu) agree with that of Kovaleva {\it et al.}\cite{Kovaleva} for LaMnO$_3$ samples all over the spectral range with a visible discrepancy only below 1.5 eV where  ellipsometric technique fails to reliably reproduce the optical response because of a crucial role of the sample morphology, surface roughness,  and depolarization effects. Namely these effects rather than intrinsic electronic structure are believed to be  responsible for a sizeable scatter of $\varepsilon_{2ab}$ data for different RMnO$_3$ samples (R=Pr, Nd, Sm, Eu) below 1.5 eV.  The spectral limitations affect smoothly the results in the low absorption range, but not the positions and the strength of intensive absorption bands. In other words, these limitations are believed not to affect the main result of our paper that is the assignment of \emph{p-d} and \emph{d-d} CT transitions in the spectral range 1.5-5 eV.

 \section{Discussion}
\subsection{Optical response of RMnO$_3$ as an interplay of \emph{p-d} and \emph{d-d} CT transitions}

In Fig.\,\ref{fig6} we have presented the results of a semi-quantitative analysis of the spectra in the energy range covered by our experiment based on a dispersion analysis of the complex dielectric function $\varepsilon $, which  was  fitted by a set of 8 Lorentzian oscillators. To  reproduce the  tails of the high-energy bands  we have  introduced an  additional optical band peaked  at around 8.7 eV ($\varepsilon_{ab}$) or 8.2 eV ($\varepsilon_{c}$). It is worth noting, that the Lorentzian fitting should be made with a great care because of several points. First of all, the Lorentzian function  is a clear oversimplification for complex line-shapes  resulting  from an interplay of the electron-lattice interaction and excitonic band effects. Second, it is worth noting the increased uncertainties both of the measurement and, accordingly, of  the  analysis close to the low- and high-energy cutoffs of our experiment. It should be noted that  the results of the Lorentzian fitting for the $\varepsilon _1$ near 2 eV optical feature rather than $\varepsilon _2$ are quite robust to the assumptions related with the transitions beyond the energy range.
All the oscillators we made use can be sorted into three groups assigned to: i) an intensive and broad band peaked at 2.2 eV, ii) weak and narrow bands peaked at 1.3, 1.9, 2.3, 2.65 eV, forming a fine structure of 2.2 eV band, and iii) intensive bands peaked at 3.0, 3.9, 4.65, 5.3 eV, forming a broad spectral feature peaked at around 4-5 eV. A proper Lorentzian decomposition is shown in Fig.\,\ref{fig6}.

Both the spectral weight, polarization properties, and temperature behavior\cite{Quijada,Kovaleva,Kovaleva-09} of the intensive and broad 2 eV band in all the manganites investigated point to its inter-site \emph{d-d} CT character, or strictly speaking,  to the low energy HS $e_g-e_g$ CT transition. Such a conclusion is strongly supported by proper quantitative estimates\,\cite{Ahn,Kovaleva,Kovaleva-09}.

One of the most impressive manifestations of the \emph{d-d} CT origin of the 2 eV band in LaMnO$_3$ is provided by the resonance  Raman spectroscopy\,\cite{Raman-Kruger}. Indeed, the out of phase B$_{2g}$ (Pbnm system) breathing  mode\,\cite{Martin} at 611 cm$^{-1}$, which strongly modulates the intersite \emph{d-d} charge transfer, shows a single pronounced resonance at around 2 eV. It is worth noting that the authors\,\cite{Raman-Kruger} found a good agreement
between the experimental results and a theory by Allen and Perebeinos\,\cite{Allen} based on the Franck-Condon mechanism activating multiphonon Raman scattering. However, the assignment of the 2 eV absorption peak  to the intra-molecular transition due to the Frank-Condon process between the Jahn-Teller split $E_g$ levels seems to be questionable because it cannot explain the basic features of 2 eV band such as a rather strong spectral weight with a peculiar temperature dependence (see detailed discussion in Ref.\,\onlinecite{Quijada}).

As it is seen from Exp.(\ref{SWabc}) the spectral weight of the 2 eV \emph{d-d} CT   band depends on the orbital mixing angle $\Theta$, Mn-O-Mn bond angle $\theta$, and the \emph{d-d} transfer energy $\Delta$, which in turn depends on the JT distortions of MnO$_6$ octahedra (see Exp.(\ref{Delta})). Our experimental data unambiguously show that the minimal inter-site CT energy $\Delta \approx 2.2$ eV is virtually  constant for R= Nd, Pr, Sm, and Eu. Interestingly that the red shift of the band in LaMnO$_3$ ($\Delta _{La} =$ 2.0 eV) is due to a sizeably lower magnitude of the JT energy $\epsilon _{JT}\propto T_{JT}$ (see Ref.\onlinecite{Good2}) as compared with other manganites studied  ($\Delta _{Pr,Nd,Sm,Eu} =$ 2.2 eV), furthermore, this can provide a sizeable 20\% contribution to the spectral weight enhancement, while the effect of  $\Theta$ and  $\theta$ variation is restricted by $\sim$10\%.
This result  agrees with theoretical predictions based on the structural dependence of the SW and also supports the inter-site \emph{d-d} CT assignment of the 2 eV band. 

Lorentzian decomposition of 2 eV feature in RMnO$_3$ with assignement of all the peaks is shown in Fig.\ref{fig6}.
The intensity and energy position of four narrow bands forming the fine structure of 2 eV feature agree surprizingly well with the theoretical predictions made earlier by one of the authors (see Ref.\onlinecite{Moskvin-Mn}) for the forbidden $t_{1g}(\pi)\rightarrow e_g$ and weakly-allowed  $t_{2u}(\pi)\rightarrow e_g$ one-center \emph{p-d} CT transitions splitted by a near tetragonal crystal field. It is worth noting that a close similarity of tetragonal distortions for MnO$_6$ octahedra in all the manganites under consideration\,\cite{Good2} does explain a nearly equal position of relatively weak and narrow bands peaked  for all the manganites at 1.3, 1.9, 2.3, 2.7\,eV ($\varepsilon_{ab}$).
  Weak non-tetragonal distortions of MnO$_6$ octahedra are seemingly visible for $t_{1g}(\pi)\rightarrow e_g$ transition. Indeed, the $ab$-response is well fitted by two Lorentzians peaked at 1.3 and 1.9 eV, while $c$-response needs two Lorentzians peaked at 1.3 and 1.7 eV, respectively. It is worth noting that the energy splitting of nonbonding oxygen $t_{1g}(\pi),t_{2u}(\pi)$ orbitals ($\approx$0.5 and $\approx$\,0.4 eV, respectively) appears to be slightly smaller than that of $e_g$ orbital ($\Delta _{JT}\approx$ 0.7 eV). The relative sharpness of all the weak \emph{p-d} CT bands agrees with a localized character of  nonbonding oxygen O 2\emph{p}$\pi$  orbitals.
 The most intensive and broad high-energy band modeled by four Lorentzians peaked at 3.0, 3.9, 4.65, and 5.3\,eV can be unambiguously assigned to the one-center dipole-allowed \emph{p-d} CT transitions: $t_{1u}(\pi)-e_{g}$($\pi -\sigma$),   $t_{2u}(\pi)-t_{2g}$($\pi -\pi$), $t_{1u}(\sigma)-e_{g}$($\sigma -\sigma$) transitions, respectively, with a hardly resolved low-symmetry splitting for the most intensive and broad bands. The composite high-energy \emph{p-d} CT band is markedly revealed by the resonance  Raman spectroscopy in LaMnO$_3$\,\cite{Raman-Kruger}. One observe pronounced resonances at around 4-5 eV both for the A$_g$ Jahn-Teller stretching and A$_g$ out-of-phase rotation   modes of the oxygen octahedra at 496 cm$^{-1}$ and 284 cm$^{-1}$, respectively, and also for the in phase symmetric B$_{1g}$ stretching  mode at 655 cm$^{-1}$. Interestingly that the same Raman modes resonate also at around 2 eV that supports our conclusion about the composite \emph{d-d}/\emph{p-d} structure of the low-energy CT band. It is worth noting that in LaMnO$_3$ we deal with a near degeneracy of the \emph{d-d} CT transition and one of the \emph{p-d} CT transitions at 1.9-2.0 eV. It means a strong coupling/mixing of two types of transitions.
 
Concluding the subsection we would like to comment an assignment of \emph{p-d} and \emph{d-d} CT transitions in LaMnO$_3$ reported in recent papers by Kovaleva {\it et al}\,\cite{Kovaleva,Kovaleva-09}. As it is clearly seen in Fig.\,\ref{fig6}(see also Figs.8 and 9 in Ref.\,\onlinecite{Kovaleva-09}) the optical response of LaMnO$_3$ in contrast to RMnO$_3$ (R=Pr, Nd, Sm, Eu) does not distinctly reveal fine structure effects. Neverthless, the authors\,\cite{Kovaleva,Kovaleva-09} made use of a Lorentzian decomposition of the complex dielectric response $\varepsilon _b$ and $\varepsilon _c$, measured by  ellipsometric technique in the spectral range 1.2-6.0 eV at the temperatures varied between 20 and 300\,K.  They did  consider only one high-energy strongly dipole-allowed \emph{p-d} CT transition at 4.7 eV, while  other optical bands observed at around 2.0, 2.4, 2.7, 3.8, 4.4, 4.6, 5.7 eV were assigned to different HS and LS inter-site \emph{d-d} CT transitions. The assignment was based on the temperature dependence of the partial and total spectral weights, and a parameterization of the energy spectrum of final d$^3$-d$^5$ configuration. However, the argumentation is believed to be questionable because of several unacceptable assumptions and inconsistencies. First, the authors neglect a number of one-center \emph{p-d} CT transitions to be precursors of the most intensive high-energy dipole-allowed \emph{p-d} CT transitions peaked at around 4.7 eV. As we have shown above these include the low-energy nominally dipole-forbidden, however, phonon-assisted  $t_{1g}(\pi)-e_g$ transition centered at around 1.7 eV, the weak    dipole-allowed  $t_{2u}(\pi)-e_g$ transition centered at around 2.5 eV, and a more intensive  dipole-allowed  $t_{1u}(\pi)-e_g$ transition centered at around 3.5 eV. All the \emph{p-d} CT bands are characterized by a sizeable temperature dependence of the line-shape and spectral weight, both due to electron-lattice and spin-spin correlations, with a pronounced T-dependent SW transfer between dipole-allowed and dipole-forbidden bands. The situation becomes more involved, if to take into account the interaction/mixing between \emph{p-d} and \emph{d-d} transitions.

Second, the analysis\,\cite{Kovaleva,Kovaleva-09} of the electron structure and energy spectrum of the d$^3$-d$^5$ configuration to be a final one for the d($e_g$)-d($e_g$) charge transfer should be revisited in several points, in particular, as regards the assignment of the LS transitions. The fact is that the gap in between the low-energy HS and LS d($e_g$)-d($e_g$) CT transitions $\Delta_{HS-LS}=10\,B+5\,C\approx 3$\,eV, i.e. markedly bigger than 2.3 eV suggested in Ref.\,\onlinecite{Kovaleva-09}. The authors have unreasonably  turned down  the ${}^{4}E_g({}^{4}D)$ term of the d$^5$ configuration while the both ${}^{4}E_g({}^{4}G)$ and ${}^{4}E_g({}^{4}D)$ terms are almost equally involved in the d($e_g$)-d($e_g$) charge transfer under consideration. As a result, both the assignment of the LS \emph{d-d} CT transitions and quantitative estimates of such parameters as the Hund exchange integral $J_H$, spectral weights SW$_{LS}$ obtained in Ref.\,\onlinecite{Kovaleva-09} seem to be questionable ones.  

Third, the authors address three bands at 2.0, 2.4, and 2.7 eV to form a three-subband structure of the 2 eV \emph{d-d} CT band. However, they cannot explain the origin of this splitting. Nevertheless,  their quantitative analysis  implies $E_1=2.0$\,eV (not 2.4 or 2.7 eV!) to be an energy of the HS d($e_g$)-d($e_g$) CT transition.

From the other hand, the excellent experimental data by Kovaleva {\it et al}\,\cite{Kovaleva,Kovaleva-09} for  LaMnO$_3$ are beyond all doubt. However, these  support our assignment  of \emph{p-d} and \emph{d-d} CT transitions in perovskite manganites and challenge a detailed analysis of the temperature effects with taking into account a sizeable contribution of the electron-phonon coupling, comparable spin-dependent terms in the spectral weight for the both \emph{p-d} and \emph{d-d} CT transitions, and coupling/mixing of the both types of CT transitions.


 \subsection{Parent perovskite manganites: Mott-Hubbard or charge transfer insulators?}
Above we have provided an experimental evidence of a dual nature of charge gap states in perovskite manganites to be a superposition of inter-site \emph{d-d} and one-center \emph{p-d} CT transitions, a remarkable observation, which to the best of our knowledge has not been previously reported for the perovskite manganites. This implies a revisit of a standard ZSA classification scheme usually applied for these compounds.
 
Zaanen, Sawatzky, and Allen (ZSA) \cite{ZSA} have pointed out that all the insulating 3\emph{d}-electron systems can be classified into two  categories: the Mott-Hubbard  and charge transfer  insulators, depending on the nature of  the gap. In the former the gap is formed by the transition between the lower and upper Hubbard bands ($U<\Delta$), while in the latter it is formed by the CT transitions between upper filled ligand and upper Hubbard bands ($U>\Delta$). 
Actually in both cases we deal with CT transitions, however, in terms of a classification  introduced by A. Moskvin {\it et al}.\,\cite{Moskvin1,Moskvin2} these correspond to two-center (d$_1$-d$_2$-) and one-center (pd-) CT excitons, respectively. 
Usually one considers that the gap  is of CT \emph{p-d} type ($U>\Delta$)  for heavier transition metals (Cu, Ni, ...), while it is of \emph{d-d} type ($U<\Delta$) for the light transition metals (Ti, V, ...). 
However, actual situation can be rather intricate. For instance,
parent cuprates such as CuO, La$_2$CuO$_4$, Sr$_2$CuO$_3$, Sr$_2$CuO$_2$Cl$_2$ are usually addressed to be Mott-Hubbard insulators, however, the detailed analysis of recent EELS data reveals  its dual  nature because of the  gap  described by a superposition of one- and two-center CT excitons\,\cite{Moskvin1,Moskvin2}.
Simple ZSA scheme needs in a revisit as it does not take into account a number of important issues.

The parameter $\Delta _{pd}$ appears in the simple generic pd-model describing
the 3\emph{d} oxides, with the hole Hamiltonian
$$
\hat H = \sum _{i}\epsilon _{d}{\hat d}^{\dag}_{i\sigma}{\hat d}_{i\sigma}
+\sum _{j}\epsilon _{p}{\hat p}^{\dag}_{j\sigma}{\hat p}_{j\sigma}
+ \sum _{ij}(t_{pd}{\hat d}^{\dag}_{i\sigma}{\hat p}_{j\sigma}+h.c.)
$$
\begin{equation}
+ U_{d}\sum _{i}n_{id\uparrow}n_{id\downarrow}
+ U_{p}\sum _{j}n_{jp\uparrow}n_{jp\downarrow},
\label{h}
\end{equation}
where $\Delta _{pd}=\epsilon _{p}-\epsilon _{d}$, and all the terms have its
usual form.
This Hamiltonian describes fairly well the pd-hybridization effects. Depending on
the sign and numerical  value of the $\Delta _{pd}$ parameter we  come
to a rather smooth redistribution of the hole density from the cations given large positive
$\Delta _{pd}$ to anions given large negative $\Delta _{pd}$. This effect, as
it is stressed by Khomskii \cite{D.Khomskii} bears the quantitative character albeit
the properties of systems with positive and negative $\Delta _{pd}$ can differ
substantially.

However, simple Hamiltonian (\ref{h}) does not take into account a number of
important effects, including the symmetry of different d-, p-states,
crystalline field,  polarization corrections, and electron-lattice coupling, particularlly important in the case of a near degeneracy for the {\it centers of gravity} of the d- and p- manifolds. Electron correlations need in a more correct description, which must include a full set of the Slater, or Racah parameters.
 First of all, we should discriminate the (anti)bonding and nonbonding orbitals.
 For instance, in  simple octahedra like MeO$_6$ we have two sets of 3\emph{d} orbitals
 with $e_g(\sigma)$ and $t_{2g}(\pi)$ symmetry, and seven sets of O\,2\emph{p} orbitals with
 $a_{1g}(\sigma)$, $t_{1g}(\pi)$, $t_{1u}(\sigma)$, $t_{1u}(\pi)$, $t_{2u}(\pi)$,
 $e_g(\sigma)$, and $t_{2g}(\pi)$ symmetry. Only two of them, $e_g$ and $t_{2g}$ hybridize with
 3\emph{d} orbitals having the same symmetry. From the other hand, namely the nonbonding O\,2\emph{p} orbitals seemingly
 form the first electron removal state in perovskites with octahedrally
 coordinated 3\emph{d} ions\,\cite{Licht}. Thus, simple cluster model points to a need in the realistic multi-band
 description of 3\emph{d} oxides with a set of parameters
 like $\Delta _{pd}$\,\cite{Moskvin-02}. 
 The near degeneracy of the 3\emph{d} orbitals with nonbonding oxygen states
 qualitatively differs from that for two hybridized orbitals of the same
  symmetry, and can result in  a number of novel effects, in particular,
  generated by enhanced  electron-lattice coupling with local, or cooperative
low-symmetry lattice deformations. In the case of the near-degeneracy of
hybridizing 3\emph{d} and O\,2\emph{p} states the lattice deformation is governed by the active
high-symmetry breathing modes.

We should note that in many cases the charge transfer  with the lowest energy
corresponds to the dipole-forbidden transition with a very weak spectral weight,
that hampers its assignment. In general,  we should make use of
 $\Delta _{pd}$ symbol for the smallest of the O 2\emph{p}- Me 3\emph{d} charge transfer gaps, and
$\Delta _{CT}^{(\sigma ,\pi )}\geq\Delta _{pd}$ for the gap revealed in optical response due to strong dipole-allowed $\sigma -\sigma$ and $\pi -\pi$ charge transfer  transitions, respectively. 
Hence, one might say that the ZSA classification needs in detailization.

As concerns the ZSA classification for perovskite manganites the situation seems to be far from being resolved. The parent compound LaMnO$_3$ is sorted either into the charge transfer insulator\cite{Arima,Tobe} or the Mott-Hubbard insulator\cite{Quijada,Kovaleva,Kovaleva-09,Kim1,Kim2}. Interestingly, that in the study by Arima {\it et al.} \cite{Arima} LaMnO$_3$ is addressed to be located in the vicinity of the borderline where the optical gap changes from the Mott gap to the charge transfer gap that implies its dual nature. 
Our experimental data and theoretical analysis support this conjecture and evidence a dual nature of the dielectric gap in perovskite manganites RMnO$_3$, being formed by a superposition of \emph{p-d} CT transitions and inter-site \emph{d-d} CT transitions.
In fact, the parent perovskite manganites RMnO$_3$ should rather be sorted neither into the CT insulator nor the Mott-Hubbard insulator in the ZSA scheme. 

At present there is no decisive experimental technique that provides the separation of one- and two-center CT excitons and distinction between Mott-Hubbard and CT insulators. It seems, one-dimensional 3\emph{d} compounds are good candidates for such a study since the light  with the electric field
${\bf E}$ perpendicular to the chain direction does excite only
electron-hole pairs sitting on one MeO$_n$ cluster. On the other
hand, for ${\bf E}$ parallel to the chain direction, both types of
excitons (one- and two-center's) can be observed. Indeed, the polarization-dependent angle-resolved EELS
study of the 1D cuprate Sr$_2$CuO$_3$ with the corner-shared CuO$_4$
plaquettes has provided a unique opportunity  to separate both the
one- and two-center CT excitons and reveal 
the two-peak nature
of the CT gap  with the presence of nearly
degenerate two types of excitations\,\cite{Moskvin2}. In other words, similar to RMnO$_3$, insulating cuprates appear to be neither Mott-Hubbard, nor CT insulators, thus falling into {\it intermediate region} in which there are strong fluctuations between one- and two-center CT excitons. 

\section{Conclusions}


We have performed a compehensive theoretical and experimental study of optical response of parent perovskite manganites RMnO$_3$. Starting with a simple cluster model approach  we addressed both the one-center (\emph{p-d}) and two-center (\emph{d-d}) CT transitions, their polarization properties, the role played by structural parameters, orbital mixing, and spin degree of freedom. Optical complex dielectric function of single crystalline samples of RMnO$_3$ (R=La, Pr, Nd, Sm, Eu) has been measured by ellipsometric technique at room temperature in the spectral range from 1.0 to 5.0 eV for two light polarizations: $\bf E \parallel \bf c$ and $\bf E \perp \bf c$. The comparative analysis of the spectral behavior of  $\varepsilon _1$ and  $\varepsilon _2$, particularly for $\bf E \perp \bf c$ polarization,  is believed to provide a more reliable assignment of spectral features.  Our experimental data and theoretical analysis evidence a dual nature of the dielectric gap in  these manganites, being formed by a superposition of forbidden $t_{1g}(\pi)\rightarrow e_g$ (1.3 and 1.9 eV) or weak dipole allowed $t_{2u}(\pi)\rightarrow e_g$ (2.3 and 2.7 eV) \emph{p-d} CT transitions and inter-site d($e_g$)-d$e_g$) (2.0 eV in LaMnO$_3$ and 2.2 eV in other manganites) HS CT transitions, a remarkable observation, which to the best of our knowledge has not been previously reported for these materials. 
 In fact, the parent perovskite manganites RMnO$_3$ should rather be sorted neither into the CT insulator nor the Mott-Hubbard insulator in the Zaanen, Sawatzky, Allen  scheme.
We found an overall agreement between experimental spectra and theoretical predictions based on the theory of one-center \emph{p-d} CT transitions  and inter-site \emph{d-d} CT transitions. Spectral features at 3.0 and 3.8 eV are assigned to a weak dipole-allowed $t_{1u}(\pi)\rightarrow e_g$ \emph{p-d} CT transition of the $\pi - \sigma$ type with a clearly resolved low-symmetry splitting. Strong dipole-allowed \emph{p-d} CT transitions $t_{1u}(\pi)\rightarrow t_{2g}$ and $t_{1u}(\sigma)\rightarrow e_g$ of  the $\pi - \pi$ and  $\sigma - \sigma$ type, respectively, are observed as  the most intensive bands with an unresolved structure at 4.7 and 5.3 eV. Weak low-spin counterparts of the HS d($e_g$)-d$e_g$) CT transition are seemingly superimposed on these strong \emph{p-d} CT bands. 

We argue that the both dipole-allowed \emph{d-d} and \emph{p-d} CT transtions are equally sensitive to the temperature-dependent spin correlations, only with a different relative change of the optical response through the onset of the magnetic order, relatively large for the former and relatively small for the latter. 

Finally one should note that some uncertainties and discrepancies observed in optical response of different $R$MnO$_3$ samples can be related with a charge transfer instability of perovskite manganites and a trend to a phase separation.\,\cite{Moskvin-98,Moskvin-09,Moskvin-LTP,optics-PSa,optics-PSb,droplet,CIC,Mih}

 We thank R.\,V. Pisarev for fruitful
discussions. The present work was partly supported by the  RFBR under Grants No. 08-02-00633 and No. 10-02-96032 (A.S.M.), the DPS program "Physics of new 
materials and structures" (N.N.L).


\begin{thebibliography}{99}
\bibitem{Tokura}
Y.~Tokura, Rep. Prog. Phys. {\bf 69} 797 (2006).

\bibitem{Kimura}
T.~Kimura, T.~Goto, H.~Shintani, K.~Ishizaka, T.~Arima, and
Y.~Tokura, Nature  London  {\bf 426}, 55  (2003).

\bibitem{Kim2}
M.\,W.~Kim, S.\,J.~Moon, J.\,H.~Jung, J.~Yu, S.~Parashar, P.~Murugavel, J.\,H.~Lee, and T.\,W.~Noh, Phys. Rev. Lett. {\bf 96}, 247205 (2006).

\bibitem{Arima}
T.~Arima, Y.~Tokura,  J. Phys. Soc. Jap.
 {\bf 64},  2488 (1995).

\bibitem{Okimoto}
Y.~Okimoto, T.~Katsufuji, T.~Ishikawa, A.~Urushibara, T.~Arima, and Y.~Tokura,
 Phys. Rev. Lett.
{\bf 75},   109 (1995); Y.~Okimoto, T.~Katsufui, T.~Ishikawa,  T.~Arima, and Y.~Tokura, Phys. Rev. B {\bf 55}, 4206 (1997).


\bibitem{Jung}
J.\,H.~Jung, K.\,H.~Kim, D.\,J.~Eom, T.\,W.~Noh, E.\,J.~Choi, J.~Yu, Y.\,S.~Kwon, Y.~Chung, Phys. Rev. B {\bf 55}, 15489 (1997);J.\,H.~Jung, K.\,H.~Kim, T.\,W.~Noh, E.\,J.~Choi, and J.~Yu,  Phys. Rev. B {\bf 57}, R11043
(1998).

\bibitem{Takenaka}
K.~Takenaka, K.~Iida, Y.~Sawaki, S.~Sugai, Y.~Moritomo, and A.~Nakamura, J. Phys. Soc. Jap. {\bf 68}, 1828 (1999).

\bibitem{Allen}
P.B. Allen and V. Perebeinos, Phys. Rev. Lett. {\bf 83},   4828 (1999).
 
\bibitem{Quijada} 
M.\,A.~Quijada, J.\,R.~Simpson, L.~Vasiliu-Doloc, J.\,W.~Lynn, H.\,D.~Drew, Y.\,M.~Mukovskii, and S.\,G.~Karabashev,  Phys. Rev. B {\bf 64}, 224426 (2001). 

\bibitem{optics-PS}
N.\,N.~Loshkareva, Yu.\,P.~Sukhorukov, E.\,V.~Mostovshchikova, L.\,V.~Nomerovannaya, A.\,A.~Makhnev, S.\,V.~Naumov, E.\,A.~Gan'shina, I.\,K.~Rodin, A.\,S.~Moskvin, A.\,M.~Balbashov, JETP {\bf 94}, 350 (2002).
 

\bibitem{Kovaleva}
N.\,N.~Kovaleva,  A.\,V.~Boris, C.~Bernhard, A.~Kulakov, A.~Pimenov, A.\,M.~Balbashov, G.~Khaliullin, and B.~Keimer, Phys. Rev. Lett. {\bf 93}, 147204 (2004).

\bibitem{Kovaleva-09}
N.\,N.~Kovaleva,  A.\,M.~Ole\'{s}, A.\,M.~Balbashov, A.~Maljuk, D.\,N.~Argyriou, G.~Khaliullin, and B.~Keimer, arXiv:0907.5098v1.


\bibitem{Tobe}
K.~Tobe, T.~Kimura, Y.~Okimoto, Y.~Tokura, Phys. Rev. B {\bf 64}, 184421 (2001).

\bibitem{Moskvin-Mn} 
A.\,S.~Moskvin, Phys. Rev. B {\bf 65}, 205113 (2002).

\bibitem{Bastjan}
M.~Bastjan, S.\,G.~Singer, G.~Neuber, S.~Eller, N.~Aliouane, D.\,N.~Argyriou, S.\,L.~Cooper, and M.~R\"ubhausen, Phys. Rev.B {\bf 77}, 193105 (2008).

\bibitem{Lawler}
J.\,F.~Lawler, J.\,G.~Lunney, and J.\,M.\,D.~Coey.  J. Appl. Phys. Lett.
 {\bf 65}, 3017  (1994).
 
\bibitem{Kim1}
M.\,W.~Kim, P.~Murugavel, S.~Parashar, J.\,S.~Lee, and T.\,W.~Noh, New Journal of Physics {\bf 6} 156 (2004).  

\bibitem{Ahn}
K.\,H.~Ahn and A.\,J.~Millis, Phys. Rev.B {\bf 61}, 13545 (2000).

\bibitem{Ravindran}
P.~Ravindran, A.~Kjekshus, H.~Fjellvag, A.~Delin, and O.~Eriksson, Phys. Rev.B {\bf 65}, 064445 (2002).

\bibitem{Kahn}
 F.\,J.~Kahn, P.\,S.~Pershan, J.\,P.~Remeika, Phys. Rev. {\bf 186}, 891  (1969).
 
\bibitem{Moskvin-Ferro}
A.\,S.~Moskvin, A.\,V.~Zenkov, E.\,I.~Yuryeva, V.\,A.~Gubanov,
Physica B, {\bf 168},186 (1991); A.\,S.~Moskvin, A.\,V.~Zenkov,
E.\,A.~Ganshina {\it et al.}, Journal of Physics and Chemistry of
Solids, {\bf 54}, 101 (1993). 
 
 
\bibitem{Moskvin1}
 A.\,S.~Moskvin, R.~Neudert, M.~Knupfer, J.~Fink, and R.~Hayn,
 Phys. Rev. B {\bf 65}, 180512(R) (2002).

\bibitem{Moskvin2}
A.\,S.~Moskvin, J.~M\'{a}lek, M.~Knupfer, R.~Neudert, J.~Fink, R.~Hayn, S.-L.~Drechsler, N.~Motoyama, H.~Eisaki, and S.~Uchida,  Phys. Rev. Lett.  91, 037001 (2003).

\bibitem{ferrites}
R.\,V.~Pisarev,  A.\,S.~Moskvin, A.\,M.~Kalashnikova, and Th.~Rasing, Phys. Rev. B {\bf 79}, 235128 (2009).

\bibitem{Licht}
A.\,I.~Liechtenstein, A.\,S.~Moskvin, V.\,A.~Gubanov, Sov. Phys. Solid State {\bf 24}, 2049 (1982).

\bibitem{Moskvin-02}
A.\,S.~Moskvin, I.\,L.~Avvakumov,  Physica B, {\bf 322/3-4}, 371 (2002).

\bibitem{Varshalovich}
 D.\,A.~Varshalovich, A.\,N.~Moskalev, V.\,K.~Khersonskii. Quantum Theory of Angular Momentum
  (World Scientific, Singapore, 1988).
  
\bibitem{Clementi-Raimondi}
E.~Clementi and D.\,I.~Raimondi, J. Chem. Phys.  {\bf 38}, 2686 (1963);E.~Clementi, D.\,I.~Raimondi, and W.\,P.~Reinhardt, J. Chem. Phys.  {\bf 47}, 1300 (1967).

\bibitem{TMO}
C.\,N.\,R.~Rao, B.~Raveau, Transition Metal Oxides, VCH, 1995.

\bibitem{MnO}
B.~Fromme, U.~Brunokowski, and E.~Kisker, Phys. Rev. B {\bf 58}, 9783 (1998).

\bibitem{Park}
J.-H.~Park, C.\,T.~Chen, S.-W.~Cheong, W.~Bao, G.~Meigs,
V.~Chakarian, Y.\,U.~Idzerda, Phys. Rev. Lett. {\bf 76}, 4215 (1996).


\bibitem{Good2}
J.-S.~Zhou and J.\,B.~Goodenough, Phys. Rev. Lett. {\bf 96}, 247202 (2006).

\bibitem{Oles}
A.\,M.~Ole\'s, G.~Khaliullin, P.~Horsch, and L.\,F.~Feiner, Phys. Rev. B {\bf 72}, 214431 (2005).

\bibitem{Trees}
We do not consider the so-called Trees correction which results in an additional blue-shift of the LS \emph{d-d} CT transitions.

\bibitem{thesis}  
A.\,S.~Moskvin, Dr.Sci. thesis, Moscow State University, 1984.

\bibitem{Granado}  
 E.~Granado, A.~Garcia, J.\,A.~Sanjurjo, C.~Rettori, I.~Torriani, F.~Prado, R.~Sanchez, A.~Caneiro, and S.\,B.~Oseroff, Phys. Rev. B {\bf 60}, 11879  (1999).
 
\bibitem{Mukhin} 
J.~Laverdiere, S.~Jandl, A.\,A.~Mukhin, V.\,Yu.~Ivanov, V.\,G.~Ivanov, and M.\,N.~Iliev, 
Phys. Rev. B {\bf 73}, 214301  (2006).

\bibitem{DM-JETP}
A.\,S.~Moskvin, JETP, {\bf 104}, 911 (2007).
  
\bibitem{DM-PRB}
A.\,S.~Moskvin, Phys. Rev. B {\bf 75}, 054505 (2007).

\bibitem{mechanism}
A.\,S.~Moskvin, S.-L.~Drechsler, Phys. Rev. B {\bf 78}, 024102   (2008).

\bibitem{TMS}
Y.~Tanabe, T.~Moriya, S.~Sugano, Phys. Rev. Lett. {\bf 15}, 1023 (1965).

\bibitem{Balbashov}
A.\,M.~Balbashov, S.\,G.~Karabashev, Ya.\,M.~Mukovskiy, S.\,A.~Zverkov, J. Cryst. Growth {\bf 167}, 365 (1996).

\bibitem{Mukhin1} 
A.\,A.~Mukhin,  V.\,Yu.~Ivanov, V.\,D.~Travkin, and A.\,M.~Balbashov, JMMM, {\bf 226}-{\bf 230}, 1139 (2001).

\bibitem{Kadomtseva}
A.\,M.~Kadomtseva, Yu.\,F.~Popov, G.\,P.~Vorobev, V.\,Yu.~Ivanov, A.\,A.~Mukhin, A.\,M.~Balbashov, JETP Lett. {\bf 81}, 590 (2005).

\bibitem{Nomer}
L.\,V.~Nomerovannaya, A.\,A.~Makhnev, A.\,N.~Malyuk, G.\,A.~Bolotin,
G.\,L.~Shtrapenin, and A.\,N.~Ignatenkov, Fiz. Met. Metalloved. {\bf 80}, 164 (1995).

\bibitem{Raman-Kruger}
R.~Kr\"{u}ger, B.~Schulz, S.~Naler, R.~Rauer, D.~Budelmann, J.~B\"{a}ckstr\"{o}m, K.\,H.~Kim, S-W.~Cheong, V.~Perebeinos, and M.~R\"{u}bhausen,   Phys.  Rev.  Lett.  {\bf 92},  097203 (2004).

\bibitem{Martin}
L.~Mart\'{i}n-Carr\'{o}n and A.~de Andr\'{e}s, Eur. Phys. J. B {\bf 22}, 11 (2001).



\bibitem{ZSA}
J.~Zaanen, G.\,A.~Sawatzky, and J.\,W.~Allen, Phys. Rev. Lett. {\bf 55}, 418
(1985).

\bibitem{D.Khomskii}
D.~Khomskii, Lith. Phys. J. {\bf 37}, 65 (1997); arXiv:cond-mat/0101164.

\bibitem{Moskvin-98}
A.\,S.~Moskvin, Physica B, {\bf 252}, 186 (1998).

\bibitem{Moskvin-09} A.\,S.~Moskvin, Phys. Rev. B {\bf 79}, 115102 (2009).

\bibitem{Moskvin-LTP}
A.\,S.~Moskvin,  Low Temp. Phys. {\bf 33}, 234 (2007).


\bibitem{optics-PSa}
A.\,S.~Moskvin, E.\,V.~Zenkov, Yu.\,D.~Panov, N.\,N.~Loshkareva, Yu.\,P.~Sukhorukov, E.\,V.~Mostovshchikova,  Physics of the Solid State, {\bf 44}, 1519  (2002).

\bibitem{optics-PSb}
Yu.\,P.~Sukhorukov, N.\,N.~Loshkareva, E.\,A.~Gan'shina, E.\,V.~Mostovshchikova, I.\,K.~Rodin, A.\,R.~Kaul, O.\,Yu.~Gorbenko, A.\,A.~Bosak, A.\,S.~Moskvin, and E.\,V.~Zenkov, JETP, {\bf 96}, 257 (2003).


\bibitem{droplet} 
E.\,V.~Mostovshchikova, N.\,G.~Bebenin, and N.\,N.~Loshkareva, Phys. Rev. B {\bf 70}, 012406  (2004). 

\bibitem{CIC}
N.\,N.~Loshkareva, Yu.\,P.~Sukhorukov, E.\,A.~Neifel'd, V.\,E.~Arkhipov, A.\,V.~Korolev, V.\,S.~Gaviko, E.\,V.~Panfilova, V.\,P.~Dyakina, Ya.\,M.~Mukovskii, and D.\,A.~Shulyatev, JETP {\bf 90}, 389  (2000).

\bibitem{Mih}
T.~Mertelj, D.~Kuscer, M.~Kosec, and D.~Mihailovic, Phys. Rev. B {\bf 61}, 15102 (2000-II).
  

\end{thebibliography}
\end{document}